\documentclass[11pt,letterpaper]{article}
\pdfoutput=1

\usepackage{jheppub}
\usepackage{color}
\usepackage{graphicx}
\usepackage{wrapfig}
\usepackage{slashed}

\usepackage{cancel}

\usepackage{verbatim}
\usepackage{amsmath}
\usepackage{amssymb}
\usepackage{gensymb}
\usepackage{subfig}
\usepackage{url}
\usepackage{bbold}
\usepackage{xspace}

\usepackage{multirow}
\usepackage{threeparttable}
\usepackage{paralist}

\usepackage{color}
\definecolor{darkgreen}{rgb}{0,0.5,0}

\newcommand{\GeV}{\text{GeV}}
\newcommand{\TeV}{\text{TeV}}

\newcommand{\be}{\begin{equation}}
\newcommand{\ee}{\end{equation}}

\newcommand{\bm}[1]{\boldsymbol{#1}}

\def\bea{\begin{eqnarray}}
\def\eea{\end{eqnarray}}

\usepackage{accents}
\newlength{\dhatheight}

\newcommand{\lsim}{\lesssim}
\newcommand{\gtap}{\gtrsim}
\newcommand{\ltap}{\lesssim}
\newcommand{\eg}{{\it e.g.}}
\newcommand{\ie}{{\it i.e.}}

\newcommand{\met}{\slashed{E}_{T}}

\usepackage{color}

\newcommand{\bp}{B^\prime}
\newcommand{\zp}{Z^\prime}
\usepackage[usenames,dvipsnames]{xcolor}

\begin{document}
\begin{flushright}
FERMILAB-PUB-15-353-T
\end{flushright}
\title{
Exotic Decays of Heavy $B$ quarks}

\author[a]{Patrick J. Fox}

\affiliation[a]{Theoretical Physics Department, Fermilab, Batavia, Illinois 60510, USA}

\author[b]{and David Tucker-Smith}

\affiliation[b]{Department of Physics, Williams College, Williamstown, MA 01267}

\emailAdd{pjfox@fnal.gov}
\emailAdd{dtuckers@williams.edu}

\date{\today}


\arxivnumber{}

\abstract{Heavy vector-like quarks of charge $-1/3$, $B$, have been searched for at the LHC through the decays $B\rightarrow bZ,\, bh,\,tW$.  In models where the $B$ quark also carries charge under a new gauge group, new decay channels may dominate.  We focus on the case where the $B$ is charged under a $U(1)^\prime$ and describe simple models where the dominant decay mode is $B\rightarrow bZ^\prime\rightarrow b (b\bar{b})$.  With the inclusion of dark matter such models can explain the excess of gamma rays from the Galactic center.
We develop a search strategy for this decay chain and estimate that with integrated luminosity of 300 fb$^{-1}$ the LHC will have the potential to discover both the $B$ and the $\zp$ for $B$ quarks with mass below $\sim 1.6$ TeV, for a broad range of $\zp$ masses.  A high-luminosity run can extend this reach to $2$ TeV.
}

\maketitle


\newpage
\section{Introduction}

Massive vector-like quarks exist in many extensions of the Standard Model (SM), \eg\ extra-dimensional models (both warped and flat), little Higgs theories, and composite Higgs models, and they are being actively searched for at the LHC. Because these massive states are vector-like they need not have the same SM quantum numbers as states in the SM, but in many instances they do.  We focus on that case here.  In particular, we consider massive quarks, $B$, that have the same SM charges as the right-handed bottom quark.   

These new particles can be produced through their QCD couplings and are presently searched for through the decays $B \rightarrow h b /Zb/Wt$ \cite{Khachatryan:2015gza,Aad:2015mba,Aad:2014efa};
 similarly, heavy top partners are searched for in decays $T \rightarrow h t/Z t/Wb$ \cite{Chatrchyan:2013uxa,CMS-PAS-B2G-12-017,Aad:2014efa,TheATLAScollaboration:2013sha,ATLAS:2013ima}.  The present bounds on the $B$ mass vary from $\sim 750$ GeV if the decay is purely to $Z b$, to $\sim 900$ GeV if the decay is purely to $hb$.  The bound is $\sim 790$ GeV in the Goldstone limit where the branching ratios are $B(B\rightarrow Zb):B(B\rightarrow Wt):B(B\rightarrow hb)=1:2:1$.  The bounds can be weakened if the $B$ quark decays to alternative final states.  
In this paper, we devise an LHC search strategy appropriate for one such exotic decay and estimate its potential sensitivity.

\begin{figure}[t] 
   \centering
   \includegraphics[width=0.5\textwidth]{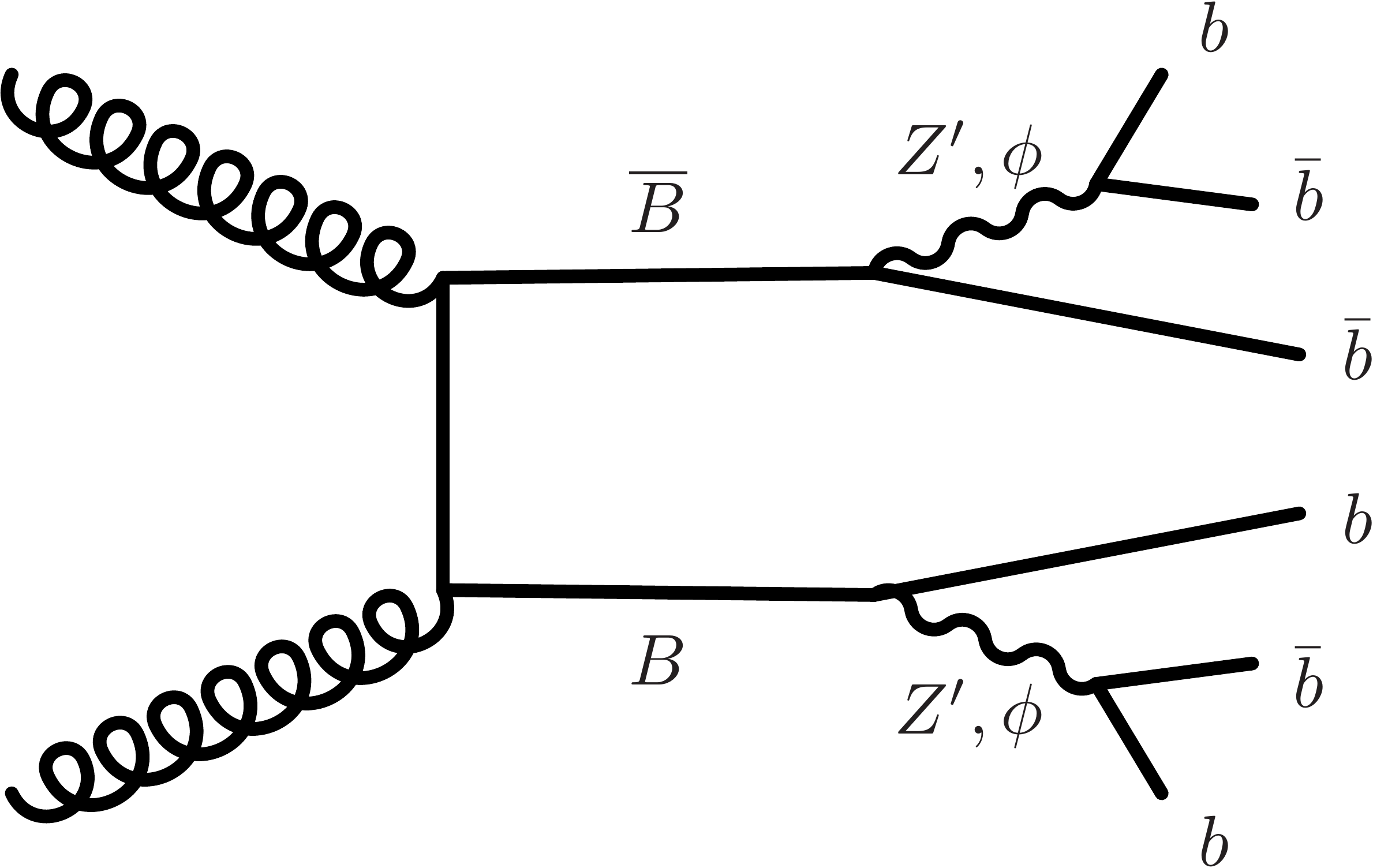} 
\caption{The $B\overline{B}$ production and decay process that is the primary focus of our analysis.}
\label{fig:BBFeynmandiag}
\end{figure}

The $B$ quark can be part of a larger extension of the SM and in particular could be charged under additional gauge groups. 
Here we consider a simple extension where the $B$ quark, which mixes with the SM $b$ quark, carries an additional $U(1)$ charge.  Such a scenario has a simple realisation within the context of ``Effective $\zp$ models" \cite{Fox:2011qd}.  These models introduce, in addition to the massive vector-like quark, a new $U(1)'$ gauge group and a scalar to break it.  Although we  focus on the case where only the vector $\zp$ is lighter than the $B$, our collider analysis will be effective provided that one or both of the $Z'$ and the scalar $\phi$ are lighter than the $B$.  In Section \ref{sec:themodel} we describe in more detail the particle content, parameter space, and phenomenology of this class of models.  We demonstrate that it is natural for the new decay chain $\bp\rightarrow b\zp\rightarrow b (b\bar{b})$, shown in Figure \ref{fig:BBFeynmandiag}, to dominate over the modes that are currently being searched for.  We also outline other interesting final states, involving SM bosons, leptons or missing energy, that can occur in some regions of parameter space and which are also interesting to search for at the LHC.

There may be other states charged under the $U(1)^\prime$, and if any are stable and electrically neutral they can be a dark matter (DM) candidate.  The annihilation products of such a DM candidate would be rich in $b$ quarks.  This presents an intriguing possibility since it is well known that the excess of high energy gamma rays seen coming from the proximity of the Galactic center \cite{Goodenough:2009gk,Hooper:2010mq} can be explained by a 30 -- 50 GeV DM particle annihilating to $b\bar{b}$, or a heavier DM particle annihilating to a pair of resonances, with mass near 50 GeV, that decay to $b\bar{b}$.  Thus, there is a possible connection between an astrophysical signal in gamma rays and a collider search in multi-$b$ final states.  We will discuss the phenomenology of the model once DM is added, and we will include as one of our collider benchmarks a scenario where the $\zp$ has a mass of 50 GeV.

Having motivated $B \rightarrow (\zp/\phi) b,\,\zp/\phi\rightarrow b\bar{b}$ as a search channel for heavy $B$ quarks we propose a new search strategy at the LHC, described in detail  in Section~\ref{sec:searching}.  
The final state contains six $b$ quarks but due to the kinematics may not contain six $b$-jets.  For this reason, 
and to be conservative,
we only require three $b$-tags in each event.  To further suppress background we find it beneficial to place a cut on the total hadronic activity in the event, $H_T\equiv\sum_{\mathrm{jets}}p_T$, that scales with the $B$ mass being searched for.

To maximize our sensitivity over a broad range of $B$ and $\zp$ masses we apply three approaches to event reconstruction, which use the hardest four, five, and six jets, respectively.  A given event is subjected to all reconstruction methods for which it qualifies, {\em e.g.} if the event has six or more hard jets all three methods are applied.
Each reconstruction method first tries to form $\zp$ candidates, keeping only those pairs of candidates whose masses are within 10\% of one another.  If $Z'$ candidates are found we then attempt to form $B$ candidates by pairing $\zp$ candidates with an extra jet, and again keep only those that are within 10\% in mass.  
The six-jet analysis reconstructs $\zp$ candidates as dijet pairs, the four-jet analysis reconstructs $\zp$ candidates as single jets with sub-structure, using the $N$-subjettiness variable~\cite{Thaler:2010tr}, and the five-jet analysis reconstructs one $\zp$ candidate as a dijet system and the other as a single jet with substructure.
For signal events the distribution of $(M_{\zp},M_B)$ pairs has a clear concentration close to the expected values.  The background distribution, coming dominantly from $t\bar{t}$ and QCD multi-jet, has a different shape, allowing separation of signal and background over a broad range of masses.

In Section~\ref{sec:Results} we present our results, which show that discovery at the 5$\sigma$ level is possible for a broad range  of $M_{Z'}$, with $M_B \lsim 1250$ GeV for 30 fb$^{-1}$, with $M_B \lsim 1600$ GeV for 300 fb$^{-1}$, and with $M_B \lsim 2000$ GeV for 3000 fb$^{-1}$.  Accurately modelling  the QCD background is a fraught enterprise.
 In a full experimental analysis the background needs to be estimated from data, and we describe one approach to doing so in Section~\ref{sec:Results}.  By relaxing the number of $b$-jets required for an event to pass the cuts one can determine the expected shape of the $(M_{\zp},M_B)$ distribution for background alone.  The normalisation of the distribution can be estimated by comparing the total number of events with and without the $b$-tags, before the analysis cuts requiring $B$ and $\zp$ candidates.  We show that  this approach works well when tested out on Monte Carlo data
 and propose other sidebands that may be available to estimate the QCD background from data.

\section{An Effective $\zp$ Model}
\label{sec:themodel}

In this section we describe a particular effective $\zp$ model ~\cite{Fox:2011qd} and identify parameter space that realizes the phenomenology we wish to study. Although  we add a relatively modest number of new fields beyond those of the SM,  several new interactions are allowed and multiple new phenomena can  arise.  We introduce a pair of vector-like quarks, $(B,B^c)$, which are charged under a new $U(1)^\prime$ and also charged under the SM in a similar way to the RH bottom quark,  \ie\ $B$ has quantum numbers $(\overline{\mathbf{3}},\mathbf{1},1/3,-1)$ under $(SU(3),SU(2),U(1)_Y,U(1)^\prime)$ and $B^c$ has $(\mathbf{3},\mathbf{1},-1/3,1)$.  Because the new quarks enter as a vector-like pair, there are no issues with gauge anomalies. In addition we introduce a new complex scalar $\Phi$ that has charge $+1$ under the $U(1)^\prime$, but which is otherwise neutral.  We assume that $\Phi$ gets a vev that breaks the $U(1)'$,
\be
\langle \Phi \rangle = \frac{w}{\sqrt{2}},
\ee
leading to a mass for the $U(1)'$ gauge field, 
\be
M_{\zp} =g' w.
\ee
For the collider phenomenology that interests us, this is the minimal model.  If  there are also vector-like fermions $(\chi,\chi^c)$ that are neutral under the SM but charged under the $U(1)^\prime$,  they can provide a viable DM candidate, as we investigate below.  An analogous setup with a vector-like top quark, $T$, in place of $B$ has been considered in Ref.~\cite{Jackson:2013rqp}.

The QCD cross section for $B\overline{B}$ pair production depends only on the mass of the vector-like quarks, but the resultant final states for these pair-production events depend upon the sizes of the various possible couplings between the SM and the new sector. 
In Section~\ref{sec:interactions} we consider these couplings and the mixings they induce.  
 In Sections~\ref{sec:Bdecays}--\ref{sec:phidecays}, we study the decays of $B$,  $Z'$, and   $\phi$.  We find that the decay chain that we use for our collider studies,  $B\rightarrow b \zp\rightarrow b(\bar{b}b)$, can easily dominate, although the analysis we develop is equally effective if  $B\rightarrow b \phi\rightarrow b(\bar{b}b)$ dominates. 
We discuss  DM phenomenology in models that incorporate the $(\chi,\chi^c)$ fields in  Section~\ref{sec:darkmatter}.

\subsection{Mixing of $B$,  $Z'$, and $\phi$ with Standard Model fields}\label{sec:interactions}
\subsubsection{Quark mixing}
If the only interactions of the vector-like quarks were their gauge interactions, there would be an unbroken $\mathbb{Z}_2$ parity under which the new fermions are odd. However, the gauge symmetries of the theory allow a so-called $\Phi$-kawa interaction, $\lambda \Phi B b^c$, which breaks the $\mathbb{Z}_2$  and allows $B$ to decay.  Including this 
Lagrangian term, the  $B$ and $b$ masses arise from
\be
\mathcal{L} \supset -m_B B B^c - \lambda \Phi B b^c - y_b Q H b^c~ + h.c.
\ee
More generally, $\Phi B$ can couple to a linear combination of $d^c$, $s^c$, and $b^c$, but to be consistent with 
flavor constraints we assume that this linear combination is dominated by $b^c$.  Alternatively, we could introduce three copies of the heavy vector-like quarks that couple in a flavor symmetric fashion to the SM down-type quarks, but with a hierarchy in the masses of the heavy quarks, such that the only sizable effective coupling of the $\zp$ is to the $b$ quark.

Either way, once $\Phi$ acquires a vev it induces $B-b$ mixing.  This mixing is largest in the RH quark sector.  The mass-eigenstate RH quark fields are
\be
\tilde{b^c} =c_R b^c - s_R B^c \,, \quad  \tilde{B^c}=c_R B^c + s_R b^c~,
\ee
with the mixing angle determined by
\be
s_R = \sin\theta_R = \frac{\lambda \langle \Phi \rangle}{M_B} = \frac{\lambda w}{\sqrt{2}M_B}~.
\label{eqn:fermion_sine_relation}
\ee
Here $M_B=\sqrt{\lambda^2 \langle \Phi \rangle^2 + m_B^2}$ is the physical mass of heavier eigenstate, and we work in the approximation that the mass of the bottom quark can be neglected.

The mixing in the LH quark sector  is related to the RH mixing  by
\be
t_L = \frac{M_b}{M_B}t_R~,
\label{eqn:fermion_tan_relation}
\ee
where as above we denote the physical mass of a field $f$ by $M_f$.
One consequence of $b-B$ mixing is that the coupling $y_b$ differs numerically from the SM bottom Yukawa coupling,  $y_b^{SM}$: 
\be
y_b = \left( \frac{ c_L}{c_R}\right) \left( \frac{\sqrt{2}  M_b}{ v}\right) = \left( \frac{ c_L}{c_R}\right) y_b^{SM},
\ee
where $v\simeq 246$ GeV.  
\subsubsection{Gauge kinetic mixing}

Another renormalizable interaction allowed by the symmetries of the theory is kinetic mixing between the $U(1)^\prime$ gauge field $(b_\mu)$ and the hypercharge gauge field $(B_\mu)$,
\be 
\mathcal{L} \supset - \frac{\kappa}{2} b_{\mu\nu} B^{\mu\nu}~.
\ee
This operator allows the $\zp$ to decay to SM fields.
If this operator is absent at some high scale $\Lambda$ (for example,  this could be the scale at which $SU(2)'$ breaks to $U(1)'$), it will be generated by $B$ and $b$ loops.  Taking $M_B$ to be somewhat above the $U(1)'$ breaking scale, we can approximate the value of $\kappa$ at the scale  $M_B$ by ignoring the quark mixing, giving
\be
\kappa \simeq \frac{g_Yg'}{6 \pi^2} \log\left(\frac{\Lambda}{M_B} \right).
\label{eqn:kinmix}
\ee
Provided $\Lambda$ is not too far above $M_B$, we expect $\kappa \sim 10^{-3} - 10^{-2}$ for $g' \sim g_Y$.  Significantly smaller values of $\kappa$ are possible for smaller $g'$, or if contributions from additional states partially cancel contributions from $b$ and $B$ loops.

Working to first order in $\kappa$, we obtain diagonal kinetic terms and  mass terms  with the field redefinitions
\bea
b_\mu &\rightarrow& c_z b_\mu + s_z Z_\mu \\
A_\mu &\rightarrow& A_\mu - \kappa c_W (c_z  b_\mu + s_z Z_\mu) \\
Z_\mu &\rightarrow& (c_z +\kappa s_W s_z)Z_\mu + (\kappa s_W c_z - s_z) b_\mu,
\eea
where $s_W$ and $c_W$ are sine and cosine of the weak-mixing angle, and the mixing angle $\theta_z$ is introduced to remove mass mixing induced by the kinetic mixing.  This mass mixing is required to be small by precision studies, and for $\kappa \ll 1$ it is guaranteed to be small  unless $M_Z$ and $M_{Z'}$ are very close.  Assuming $\theta_z\ll1$ and using the leading-order result
\be
\theta_z = \frac{M_Z^2}{M_Z^2-M^2_{\zp}}s_W \kappa~,
\ee
the couplings of the $\zp$ to SM fermions can be determined from
\be
\mathcal{L}\supset
g\left[ Z_\mu -\frac{s_W \kappa M_{\zp}^2}{M_Z^2-M^2_{\zp}} b_\mu\right]J_Z^\mu \\
+ e\left[A_\mu - \kappa c_W b_\mu \right]J_{\gamma}^\mu \\
+ g^\prime \left[ b_\mu + \frac{s_W  \kappa M_Z^2}{M_Z^2-M^2_{\zp}} Z_\mu\right]J_{\zp}^\mu
\ee
to first order in $\kappa$.
In Section~\ref{sec:zprimedecays} we consider the competition between quark mixing and kinetic mixing in determining $Z'$ branching ratios.

\subsubsection{Scalar mixing}

With the addition of $\Phi$ the scalar potential is
\be
V(\Phi,H) = -m_H^2|H|^2 +\lambda_H|H|^4 -m_\Phi^2|\Phi|^2 +\lambda_\Phi|\Phi|^4 + \lambda_{H\Phi}|H|^2|\Phi|^2~.
\ee
The mixed quartic term leads to a mass mixing between the Higgs and $\phi$ fields, producing mass-eigenstate scalars
\be
\tilde{h} = c_h h - s_h \phi  \quad\quad \tilde{\phi} = c_h \phi + s_h h, 
\ee
where
\be
\sin 2\theta_h = \frac{4\lambda_{H\Phi}\frac{M_W M_{\zp}}{g g^\prime }}{M_\phi^2-M_h^2}~
\label{eqn:scalar_mixing}
\ee
determines the mixing angle.

Scalar mixing leads to corrections to the partial widths of the SM Higgs boson of the form $\Gamma\rightarrow c_h^2 \Gamma_{SM}$, with the exception of the partial width to $b$ quarks, which is also altered by the $b-B$ mixing.  At tree level we have
\be
\Gamma(h \rightarrow b \bar{b}) =\frac{3 c_R^2 (c_L c_h y_b+s_L s_h \lambda)^2}{16 \pi }M_h ~
=\;c_L^4 c_h^2 
\left(
1+\frac{1}{\sqrt{2}} \frac{s_R \; t_h}{c_L } \frac{\lambda v}{M_B}
\right)^2
\Gamma_{SM} (h \rightarrow b \bar{b}) ~.
\ee
%
In the absence of scalar mixing,  the correction factor is
\be
c_L^4 \simeq 1-2\left( t_R \frac{M_b}{M_B} \right)^2,
\ee
and the deviation from the SM result is 
 tiny due to the smallness of $M_b$.  

If the $Z'$ is light enough, scalar mixing also induces a new decay mode,
\be
\Gamma(h \rightarrow \zp\zp) = \frac{g^{\prime 2} s_h^2}{32 \pi}\frac{\sqrt{1-4\alpha_{Z'}} \left(1-4\alpha_{\zp} +12 \alpha_{\zp}^2 \right)}{\alpha_{\zp}}M_h~,
\ee
where $\alpha_{Z'} = M_{Z'}^2/M_h^2$.
This could  lead to many interesting signatures depending on how the $\zp$ decays, \eg\ $ h \rightarrow 4b$, $h\rightarrow$ invisible (if $Z'$ decays to DM), or $h\rightarrow 4\ell$ without a $Z$ resonance.
Furthermore, the Higgs may be produced in $B$ decays (as discussed in Section~\ref{sec:Bdecays}),  resulting in a final state from $B\bar{B}$ production with as many as 10 $b$'s.  Exotic Higgs decays, {\em e.g.} $h \rightarrow Z Z'$, can also be induced by kinetic mixing.  The effects of scalar and kinetic mixing on Higgs decays have been widely studied in the literature, see for example Ref.~\cite{Curtin:2013fra}.

Beyond its effects on the Higgs particle, scalar mixing also impacts $\phi$ decays.  
In Section~\ref{sec:phidecays} we consider the competition between the $\Phi$-kawa interaction and scalar mixing  in determining $\phi$ branching ratios.

\subsection{Heavy quark decays}\label{sec:Bdecays}

As discussed above, the $\lambda \Phi B b^c$ interaction term breaks the $\mathbb{Z}_2$ parity acting on the new fermions and allows the $B$ to decay.  At tree level, the possible two-body final states are $Z'b$, $Z b$, $W^- t$, $\phi b$, and $h b$.  

For decays of $B$ into a vector boson $v$ and a fermion $f$, the relevant interaction term has the form
\be
{\mathcal L} \supset  \overline{f} \gamma^\mu \! \left( \alpha_L P_L + \alpha_R P_R \right) \!B  \;v_\mu,
\ee
and  the tree-level partial width is
\begin{eqnarray}
\Gamma(B\rightarrow vf) & =  & \frac{M_B}{32 \pi} \left( (1-x_f-x_v)^2-4 x_f x_v\right)^{1/2} \\
& & 
\hspace{-1cm}
\times
\left[ 
\left(
 \left| \alpha_L \right|^2+\left| \alpha_R \right|^2 
 \right)
\left(
\frac{1+x_v-2x_f+x_v x_f-2 x_v^2+x_f^2}{x_v} 
\right)
 -6  
 \left(
  \alpha_L^* \alpha_R+  \alpha_R^* \alpha_L 
 \right)
 \sqrt{x_f}
\right].
 \nonumber
\end{eqnarray}
Here we define $x_f = M_f^2/M_B^2$ and $x_v=M_v^2/M_B^2$. Neglecting corrections induced by kinetic mixing, the relevant couplings for  $B \rightarrow Z' b$, $B \rightarrow Z b$, and $B \rightarrow W^- t$ are
\begin{align}
\alpha_L^{Z'b} = -g' c_L s_L \quad   & \quad   \alpha_R^{Z'b} = -g' c_R s_R \\
\alpha_L^{Zb} = -\frac{e \;c_L s_L}{2c_W s_W}  \quad    &  \quad  \alpha_R^{Zb} = 0 \\
\alpha_L^{Wt} = \frac{e \; s_L}{\sqrt{2} s_W}  \quad   & \quad   \alpha_R^{Wt} = 0, 
\end{align}
where $s_L$, $c_L$, $s_R$, and $c_R$ describe the mixing in the fermion sector, with the left- and right-handed mixings  related through Equation~(\ref{eqn:fermion_tan_relation}).

For decays of $B$ into a real scalar $s$ and a fermion $f$, the relevant interaction term has the form
\be
{\mathcal L} \supset  \overline{f} \left( \beta_L P_L + \beta_R P_R \right) \!B  \;s,
\ee
and the tree-level partial width is
\begin{eqnarray}
\Gamma(B\rightarrow sf) & =  & \frac{M_B}{32 \pi} \left( (1-x_f-x_s)^2-4 x_f x_s\right)^{1/2} \\
& & 
\hspace{-1cm}
\times
\left[ 
\left(
 \left| \beta_L \right|^2+\left| \beta_R \right|^2 
 \right)
\left(
1+x_f-x_s
\right)
 +2  
 \left(
  \beta_L^* \beta_R+  \beta_R^* \beta_L 
 \right)
 \sqrt{x_f x_s}
\right],
 \nonumber
\end{eqnarray}
with $x_s=M_s^2/M_B^2$. The couplings needed for 
$B \rightarrow \phi b$ and $B \rightarrow h b$ are
\begin{align}
\beta_L^{\phi b} =-\frac{c_R}{\sqrt{2}} \left( c_L c_h \lambda + s_L s_h y_b  \right)
 \quad   & \quad   
 \beta_R^{\phi b} = \frac{s_R}{\sqrt{2}} \left( s_L c_h \lambda - c_L s_h y_b  \right)  \\
\beta_L^{hb} = \frac{c_R}{\sqrt{2}} \left( c_L s_h \lambda - s_L c_h y_b  \right)  
\quad    &  \quad
  \beta_R^{h b} =- \frac{s_R}{\sqrt{2}} \left( s_L s_h \lambda + c_L c_h y_b  \right).
\end{align}
We allow for the possibility of mixing in the scalar sector, with $s_h$ and $c_h$ determined by Equation~(\ref{eqn:scalar_mixing}).

The comparison between the various $B$ partial widths simplifies if we neglect scalar mixing ($s_h \rightarrow 0$) and work to leading non-vanishing order in $(M_b/M_B)^2$.  In this approximation we find
\bea
\Gamma(B\rightarrow \zp b) &=& 
\frac{(\lambda c_R)^2}{64 \pi}\left(1-x_{\zp}\right)^2 \left(1+2 x_{\zp}\right) M_B\\
\Gamma(B\rightarrow \phi b) &=& 
\frac{(\lambda c_R)^2}{64 \pi}\left(1-x_{\phi} \right)^2 M_B\\
\Gamma(B\rightarrow Z b) &=& 
\frac{(y_b s_R)^2}{64 \pi}\left(1-x_{Z}\right)^2 \left(1+2 x_{Z}\right) M_B\\
\Gamma(B\rightarrow h b) &=& 
\frac{(y_b s_R)^2}{64 \pi}\left(1-x_{h} \right)^2 M_B\\
\Gamma(B\rightarrow W t) &=&\frac{(y_b s_R)^2}{32 \pi} \sqrt{\left(1-x_t-x_W\right)^2-4x_t x_W}\\
&&\times \left[(1-x_t)^2+ x_W \left(1+x_t-2x_W \right)\right] M_B. \nonumber
\eea
In the regime where $M_B$ is much larger than all other masses, we have
\begin{align}
\Gamma(B\rightarrow \zp b) &  \simeq  \Gamma(B\rightarrow \phi b) \\
\Gamma(B\rightarrow Z b) \simeq \Gamma(B&\rightarrow hb) \simeq \frac{1}{2} \Gamma(B\rightarrow Wt)\\
\Gamma(B\rightarrow \zp b)  \simeq &\left( \frac{\lambda  c_R}{y_b s_R} \right)^2 \Gamma(B\rightarrow Z b),
\end{align}
consistent with Goldstone equivalence.

\begin{figure}[htbp] 
   \centering
   \includegraphics[width=0.45\textwidth]{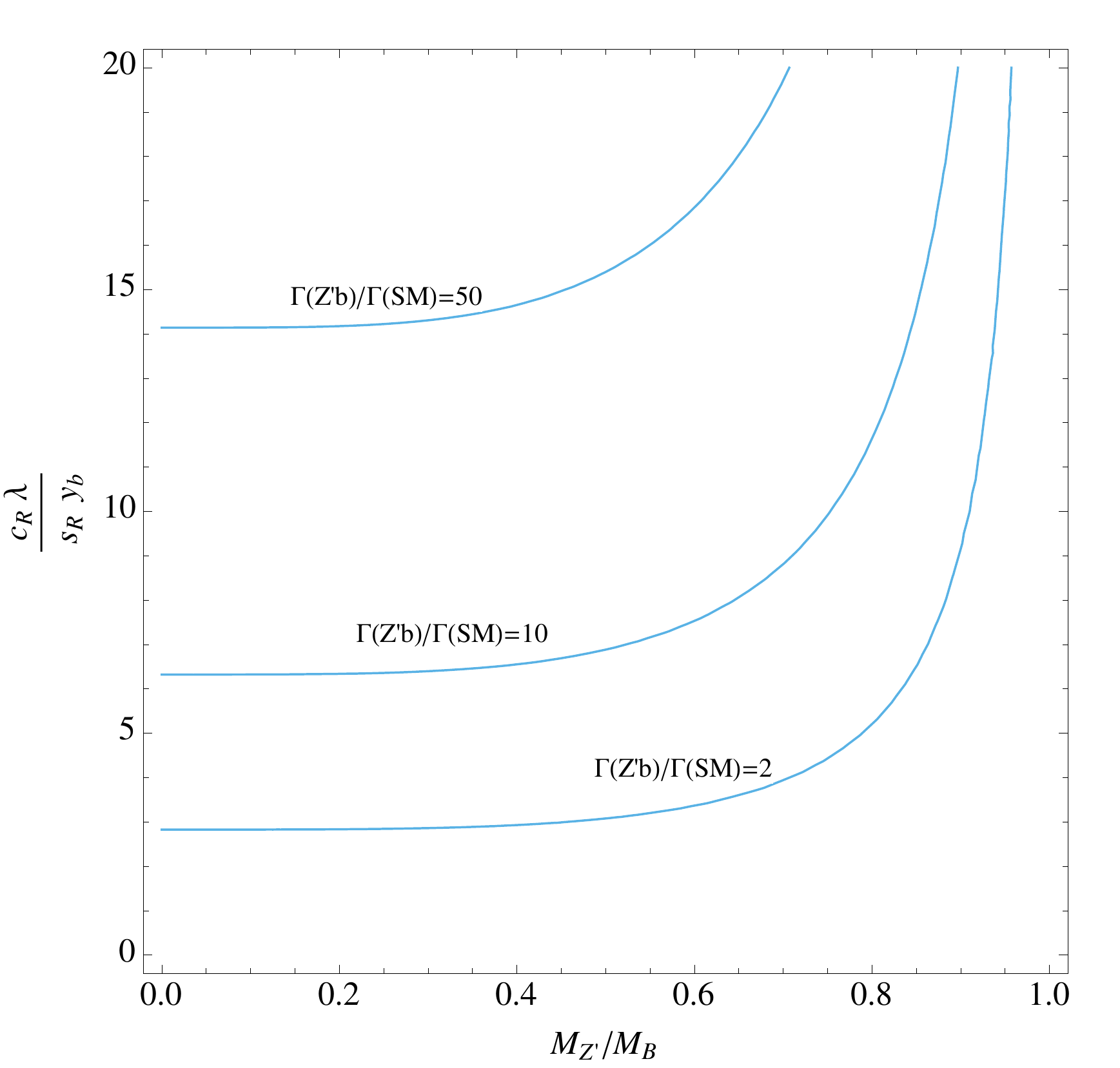} 
    \includegraphics[width=0.43\textwidth]{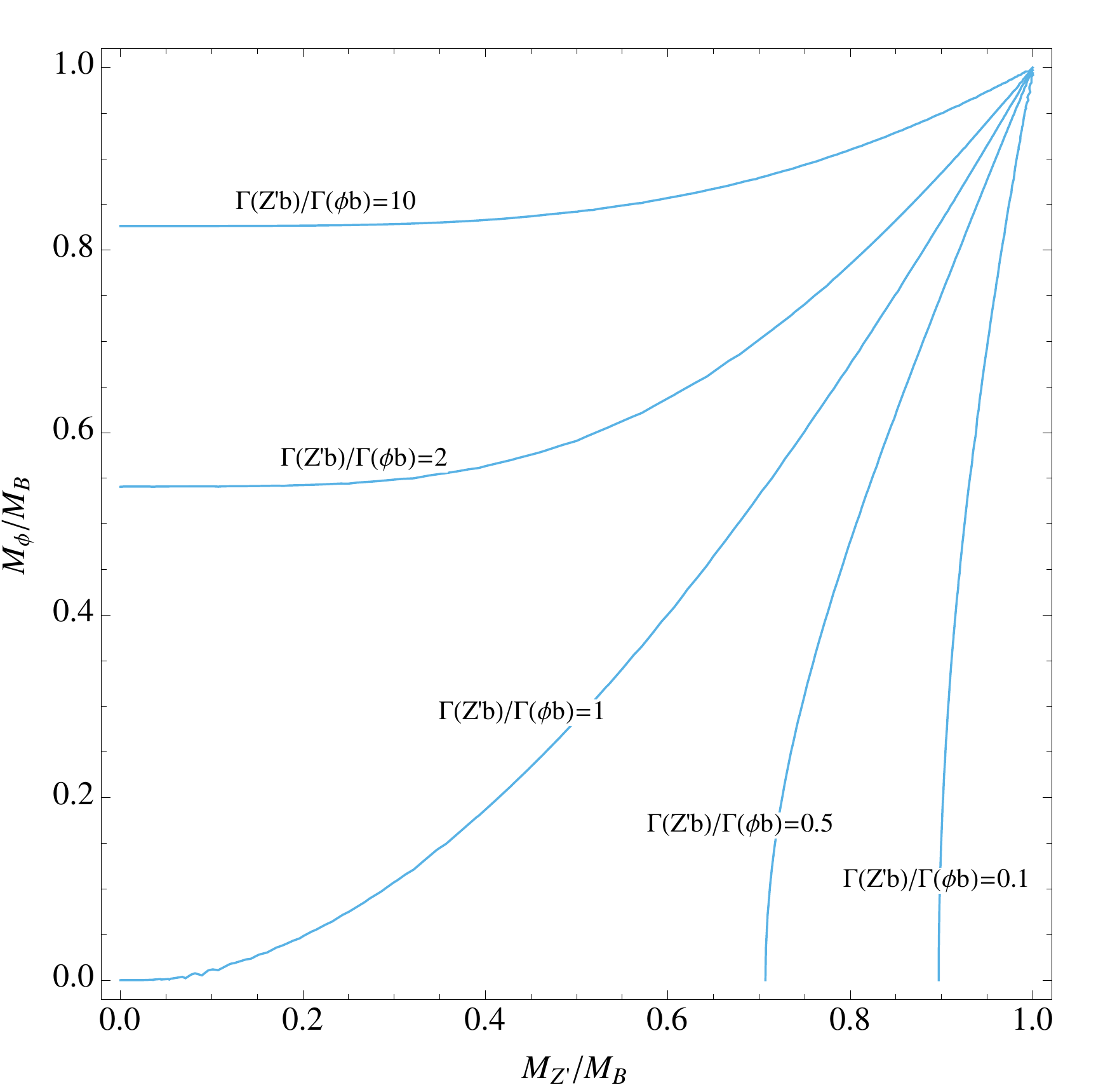}  
   \caption{{ \em Left}: Contours of $\Gamma(B\rightarrow \zp b)/\Gamma(B\rightarrow hb/Z b/W^-t)$.  We neglect the masses of all SM particles, which overestimates the partial widths into SM states.     
   { \em Right}: Contours of $\Gamma(B\rightarrow \zp b)/\Gamma(B\rightarrow \phi b)$.}
   \label{fig:Bratios}
\end{figure}

Our collider studies will focus on the  decay of $B$ to $\zp b$.  As shown in the left-hand plot of Figure~\ref{fig:Bratios},
 this decay can easily dominate over decays into SM states, due to the smallness of $y_b$. 
 In fact,  using Eqn.~(\ref{eqn:fermion_sine_relation}), the quantity appearing on the vertical axis can be rewritten as
 \be
 \frac{c_R \lambda}{s_R y_b} = \frac{c_R M_B}{y_b \langle \Phi \rangle},
 \ee
 which goes to $M_B/(y_b^{SM} \langle \Phi \rangle)$ in the $\lambda \rightarrow 0$ limit.  
It is not  therefore not necessary for $\lambda$ to be large for $B \rightarrow Z' b$ to dominate. 
Given ample phase space for the decay, 
 $B \rightarrow Z' b$ dominates over decays to SM states for small $\lambda$, unless $\langle \Phi \rangle$ is much larger than $M_B$.

  The remaining competing decay, $B \rightarrow \phi b$, can be forbidden kinematically by raising $M_\phi$ above $M_B$.  A light $Z'$ is consistent with $M_\phi > M_B$ because $g'$ can be taken to be small.  The opposite scenario is also possible:  one can have a light $\phi$ with $M_{Z'} > M_B$ if  the quartic coupling $\lambda_\phi$ is small.  In this case $B \rightarrow \phi b$  can be the dominant decay.  The right-hand plot of of Figure~\ref{fig:Bratios} shows how the ratio $\Gamma(B\rightarrow \zp b)/\Gamma(B\rightarrow \phi b)$ depends on $M_\phi$ and $M_{Z'}$ when both channels are kinematically accessible. 
 
 If  $B \rightarrow \phi b$ dominates, the results of our collider studies apply essentially unchanged, provided that $\phi$ decays dominantly to $b \overline b$ ($\phi$ decays are studied in section \ref{sec:phidecays}).   If instead both $B \rightarrow \zp b$ and $B \rightarrow \phi b$ have sizable branching ratios, the analysis we develop below is flexible enough to reconstruct both $B \overline B \rightarrow \phi \phi b \overline{b}$ events and $B \overline B \rightarrow Z' Z'  b \overline{b}$ events, even if $M_{Z'}$ and $M_\phi$ are very different.  Two invariant mass peaks at distinct values of $M_{Z'}/M_\phi$ would be found, with reduced strength compared to the case with just one dominant channel. Our analysis is not designed to reconstruct  $B \overline B \rightarrow \phi Z' b \overline{b}$ events efficiently, unless the $\zp$ and $\phi$ happen to  be close in mass.  

\subsection{$Z'$ decays}
\label{sec:zprimedecays}

At tree level, and neglecting kinetic mixing, the  potential two-body channels for $Z'$ decay are $b \overline{b}$, $b\overline{B}$, $\overline{b}B$, and $B\overline{B}$, some of which might be kinematically forbidden.  Kinetic mixing allows for decays into other fermions, including leptons, and decays to bosons.  If DM is charged under $U(1)^\prime$ and is sufficiently light, there will also be invisible decays of the $\zp$, as discussed in Section~\ref{sec:darkmatter}.

For decays of the $Z'$ into fermions $f_1 \overline{f}_2$, the interaction term 
\be
{\mathcal L} \supset  \overline{f}_1 \gamma^\mu \! \left( \rho_L P_L + \rho_R P_R \right) \!f_2 \;Z'_\mu
\ee
leads to the tree-level partial width
\begin{eqnarray}
\Gamma(Z'\rightarrow f _1 \overline{f}_2) & =  & \frac{N_c M_\zp}{48 \pi} \left( (1-y_1-y_2)^2-4 y_1 y_2\right)^{1/2} \label{eqn:genericZpffwidth}\\
& & 
\hspace{-1cm}
\times
\left[ 
\left(
 \left| \rho_L \right|^2+\left| \rho_R \right|^2 
 \right)
\left(
2-y_1-y_2 - (y_1-y_2)^2
\right)
 +6 
 \left(
  \rho_L^* \rho_R+  \rho_R^* \rho_L 
 \right)
 \sqrt{y_1 y_2}
\right],
 \nonumber
\end{eqnarray}
where $N_c$ is the fermion color multiplicity and $y_{1,2}=M_{f_{1,2}}^2/M_{\zp}^2$.

Neglecting corrections induced by kinetic mixing, the relevant couplings for  $Z' \rightarrow b \overline{b}$, $Z' \rightarrow b \overline{B}/ B \overline{b}$, and $Z' \rightarrow B \overline{B}$ are
\begin{align}
\rho_L^{b\overline{b}} = g' s_L^2 \quad   & \quad   \rho_R^{b\overline{b}} = g' s_R^2 \\
\rho_L^{B\overline{b}} = \rho_L^{b\overline{B}} =- g' s_L c_L \quad    &  \quad  \rho_R^{B\overline{b}} =\rho_R^{b\overline{B}} = -g' s_R c_R \\
\rho_L^{B\overline{B}} = g' c_L^2  \quad   & \quad   \rho_R^{B\overline{B}} = g' c_R^2.
\end{align}
Dropping terms involving $M_b^2/M_{Z'}^2$ and $M_b^2/M_{B}^2$,  we find
\bea
\Gamma(\zp \rightarrow b \bar{b}) &=& \frac{g^{\prime 2} s_R^4 }{8 \pi }M_{\zp}~\label{eq:Zprimeqm1}\\
\Gamma(\zp \rightarrow  b \bar{B}) =\Gamma(\zp \rightarrow  \bar{b} B) &=& \frac{g^{\prime 2} c_R^2 s_R^2}{16 \pi }
\left(1-y_B\right)^2 \left(2+y_B\right)M_{\zp}\\
\Gamma(\zp \rightarrow  B \bar{B}) & = &\frac{g^{\prime 2} }{8\pi}\sqrt{1-4 y_B} \left[ \left(1+c_R^4 \right)- y_B  \left(1-6 c_R^2+c_R^4 \right) \right]M_{\zp}
\label{eq:Zprimeqm2}
~.
\eea
Our collider studies will focus on scenarios with $M_{B}>M_{\zp}$, in which case $\zp \rightarrow b \bar{b}$ is the only allowed decay among those above.

Kinetic mixing modifies the $\zp$ widths given in (\ref{eq:Zprimeqm1})-(\ref{eq:Zprimeqm2}) and opens up new $Z'$ decay modes. If only kinetic mixing is present, the couplings of $Z'$ to SM fermions can be summarized as
\begin{align}
\rho_L^{f\overline{f}} = \frac{e \kappa}{c_W} \left( \frac{T^3_f - Q_f \left[1-c_W^2 y_Z \right]}{1-y_Z} \right) \quad   & \quad   \rho_R^{f\overline{f}} = -\frac{e \kappa Q_f}{c_W} \left( \frac{1-c_W^2 y_Z}{1-y_Z} \right) 
\end{align}
where we work to leading order in $\kappa$ and where $y_Z=M_Z^2/M_{\zp}^2$.  These couplings can be used with Equation~(\ref{eqn:genericZpffwidth}) to calculate the $Z'$ partial widths into SM fermions induced by kinetic mixing.   For fermions that can be approximated as massless, the result simplifies to
\begin{eqnarray}
\Gamma(Z' \rightarrow f \overline f)  & =  & \frac{N_c e^2 \kappa^2 }{24 \pi c_W^2} 
\left(\frac{
Q_f^2 \left(1-c_W^2 y_Z \right)^2
+\left(T_f^3-Q_f \left[1-c_W^2 y_Z \right] \right)^2
 }{(1-y_Z)^2} \right)
  M_\zp~.
\label{eqn:Zpffwidthkinmix}
\end{eqnarray}
Kinetic mixing also opens up decays of the $Z'$ to boson pairs, if kinematically allowed, with partial widths
\begin{align}
\Gamma(\zp \rightarrow W W)  &= \frac{e^2 \kappa ^2}{192 \pi c_W^2 }  \frac{\left(1-4 y_W\right)^{3/2} \left(1+20 y_W +12 y_W^2\right)}
{\left(1-y_Z\right)^2}M_{\zp}~\\
\Gamma(\zp\rightarrow Zh) &=\frac{e^2 \kappa^2}{192\pi c_W^2}  \sqrt{(1-y_h-y_Z)^2-4  y_h y_Z}
\left(
 \frac{ (1-y_h+y_Z)^2+8 y_Z}
{\left(1-y_Z\right)^2} 
\right)
M_{\zp}~.\label{eq:Zprimekm2}
\end{align}
If present, scalar mixing modifies $\Gamma(\zp\rightarrow Zh)$ and, for sufficiently light $\phi$, induces a partial width for $\zp\rightarrow Z\phi$ .

Large values of $\kappa$ allow for abundant $Z'$ production through its couplings to light quarks.
The $Z'$ can then decay to leptons, and LHC constraints on dilepton resonances potentially become relevant \cite{Khachatryan:2014fba}. 
For smaller $\kappa$  the $\zp$ is mainly produced through its interactions with $b$ and $B$ quarks,
but interesting leptonic signatures can still be induced by $\kappa$, \eg\ one or two dilepton resonances produced in association with $b$-jets.  

In Sections~\ref{sec:searching} and \ref{sec:Results}  we focus on the case where the dominant decay is $Z \rightarrow b\overline{b}$.
\begin{figure}[t] 
   \centering
  \includegraphics[width=0.7\textwidth]{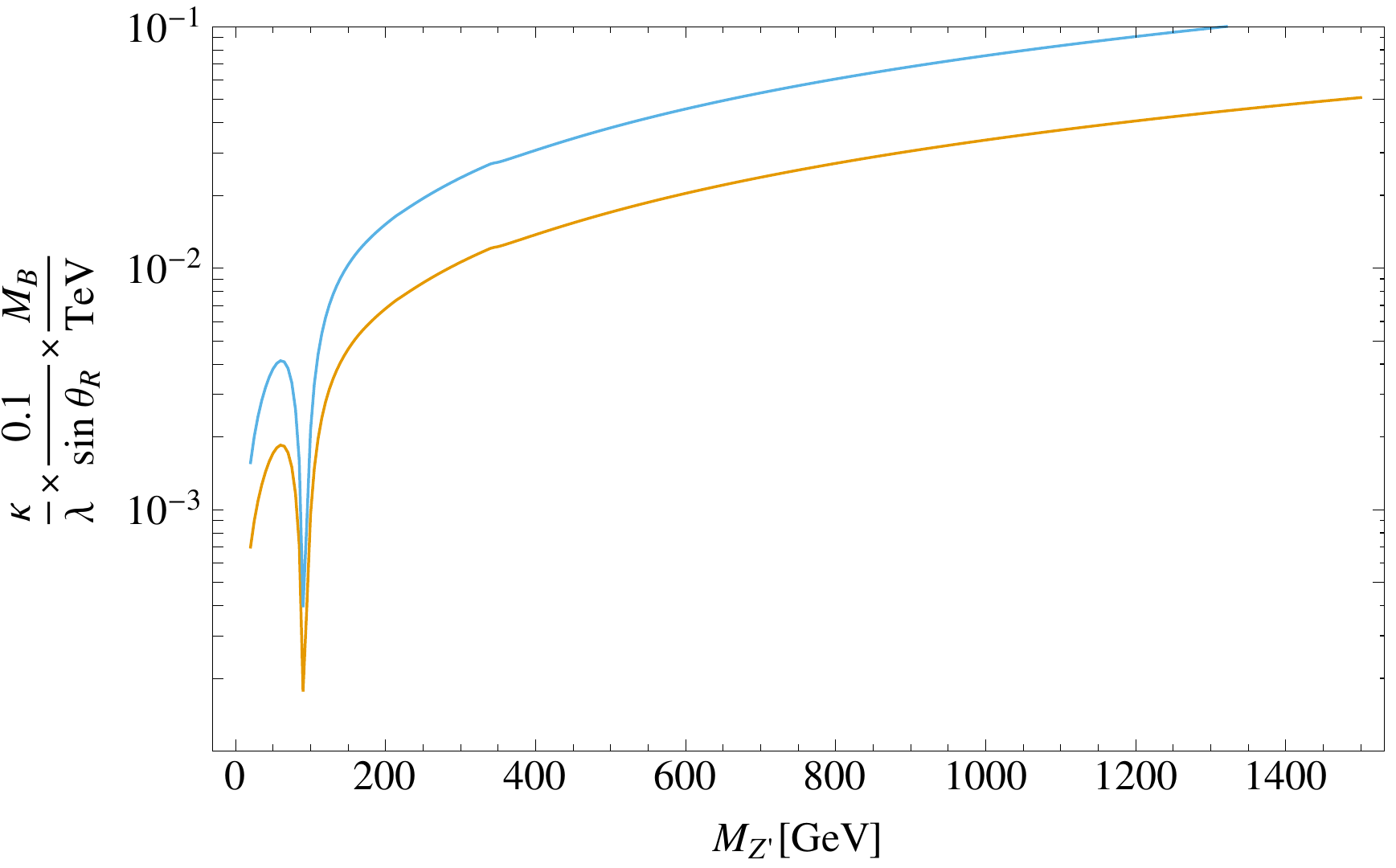}  
   \caption{In the region below the lower orange line (upper blue line), the total width of the $\zp$ induced by kinetic mixing (in the absence of quark mixing) is less than $10\%\; (50\%)$ of the partial width into $b \overline{b}$ induced by quark mixing, given by Eqn.~(\ref{eq:Zprimeqm1}).}
   \label{fig:kinmix_vs_quarkmix}
\end{figure}
To estimate what values of $\kappa$ are consistent with this scenario, we consider the ratio
\be
R_{Z'} \equiv \frac{\Gamma(Z') \text{ with only kinetic mixing turned on}}{\Gamma(Z' \rightarrow b{\overline b})\text{ with only quark mixing turned on}},
\ee
which depends on $M_{\zp}$ and on the quantity
\be
\frac{\kappa}{g' s_R^2} =\sqrt{2} \left(  \frac{\kappa M_B}{\lambda s_R M_{\zp}} \right)~.
\ee
If we require $R_{Z'}$ to be small, we get the relatively weak constraints on $\kappa$ shown in Figure~\ref{fig:kinmix_vs_quarkmix}.  Taking  $M_{B}=1\ \TeV$, $\lambda=1$, and $\sin \theta_R = 0.5$, $R_{Z'}<0.1$ implies $\kappa< 9\times 10^{-3}$ for $M_{\zp}=50\ \GeV$ and $\kappa< 9\times 10^{-2}$ for $M_{\zp}=500\ \GeV$.

\subsection{$\phi$ decays}\label{sec:phidecays}

In our discussion of $\phi$ decays we will consider the effects of scalar mixing, but we will neglect kinetic mixing.  If the scalar mixing vanishes, then at tree level, the potential two-body channels for $\phi$ decay are $Z' Z'$, $b\overline{b}$, $B \overline{B}$, $b\overline{B}$, and $B\overline{b}$.  Because we are mainly interested in how $\phi$ will decay if it happens to be produced in $B$ and $\overline{B}$ decays, we will take $M_\phi < M_B$ for this section, kinematically forbidding  decays to $B \overline{B}$, $b\overline{B}$, and $B\overline{b}$.

Scalar mixing allows the $\phi$ to acquire the decay channels of the SM Higgs.
For any decay channel $X$ open to a SM Higgs of mass $M_\phi$, excluding channels involving $b$ quarks, we have
\be
\Gamma (\phi \rightarrow X) = s_h^2 \Gamma (h \rightarrow X)|_{M_h \rightarrow M_\phi}.
\ee
The decay width to $b\overline{b}$ depends on the quark mixing.  Working to leading order in $M_b$, the tree-level width is
\be
\Gamma(\phi \rightarrow b \bar{b}) = \frac{3 c_R^2  ( s_h y_b-c_h s_L \lambda)^2}{16 \pi }M_\phi
=\; \frac{3 }{8 \pi}   \left( \frac{M_b}{v} \right)^2
 \left(   s_h - c_h s_R^2 \frac{v}{w} \right)^2
M_\phi.
\ee
The remaining two-body, tree-level partial widths are
\bea
\Gamma(\phi \rightarrow h h) &=& \frac{s_h^2 c_h^2}{32 \pi} \sqrt{1-4 z_h } (1+2 z_h)^2 \left( \frac{s_h}{w} + \frac{c_h}{v}\right)^2 M_\phi^3\,\\
\Gamma(\phi \rightarrow \zp\zp) &=& \frac{g^{\prime 2} c_h^2}{32 \pi}\frac{\sqrt{1-4z_{Z'}} \left(1-4z_{\zp} +12 z_{\zp}^2 \right)}{z_{\zp}}M_\phi~,
\eea
where $z_h = M_h^2/M_\phi^2$ and $z_{\zp} = M_{\zp}^2/M_\phi^2$.  
\begin{figure}[htbp] 
   \centering
    \includegraphics[width=0.4\textwidth]{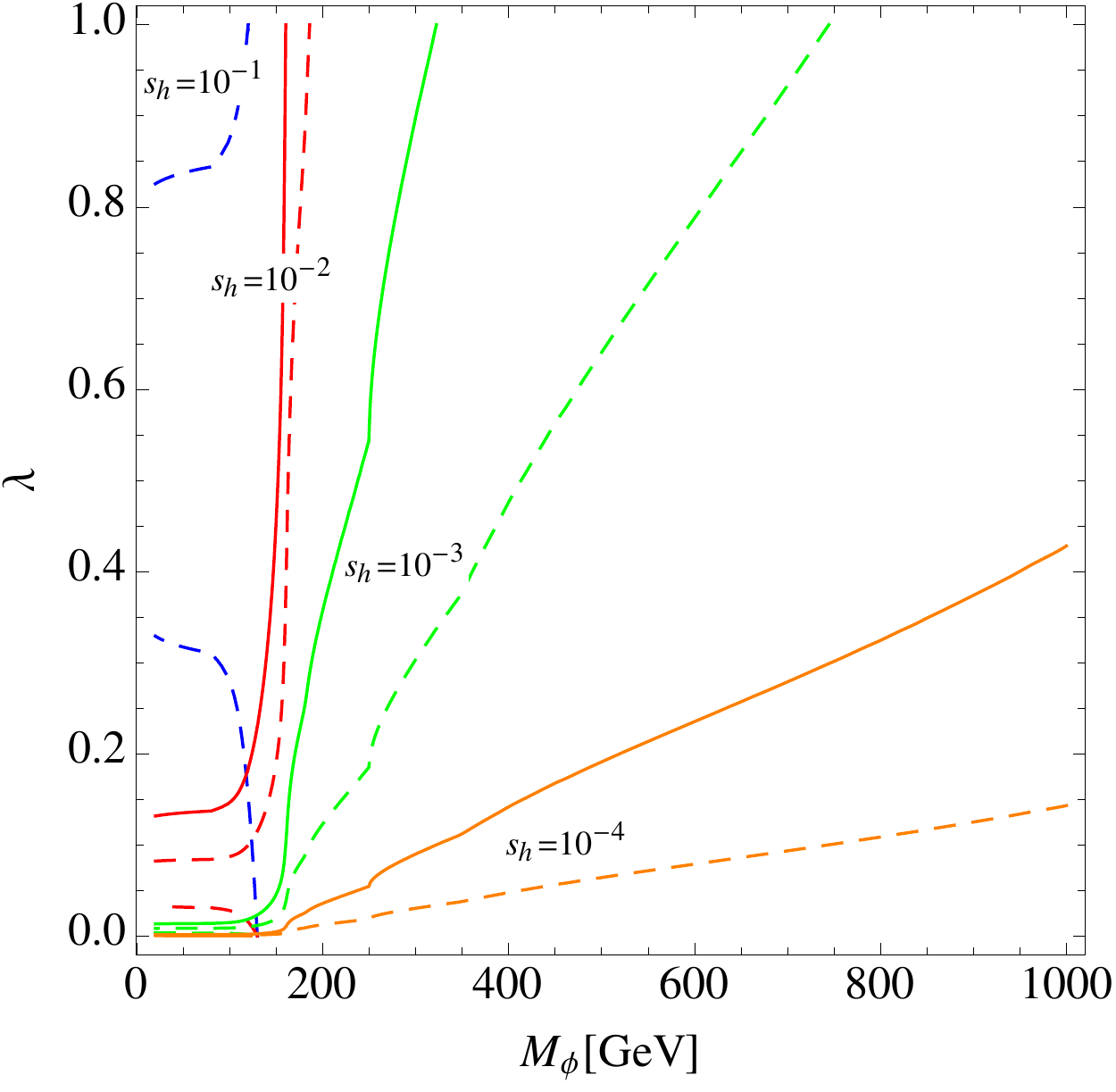}  
      \includegraphics[width=0.4\textwidth]{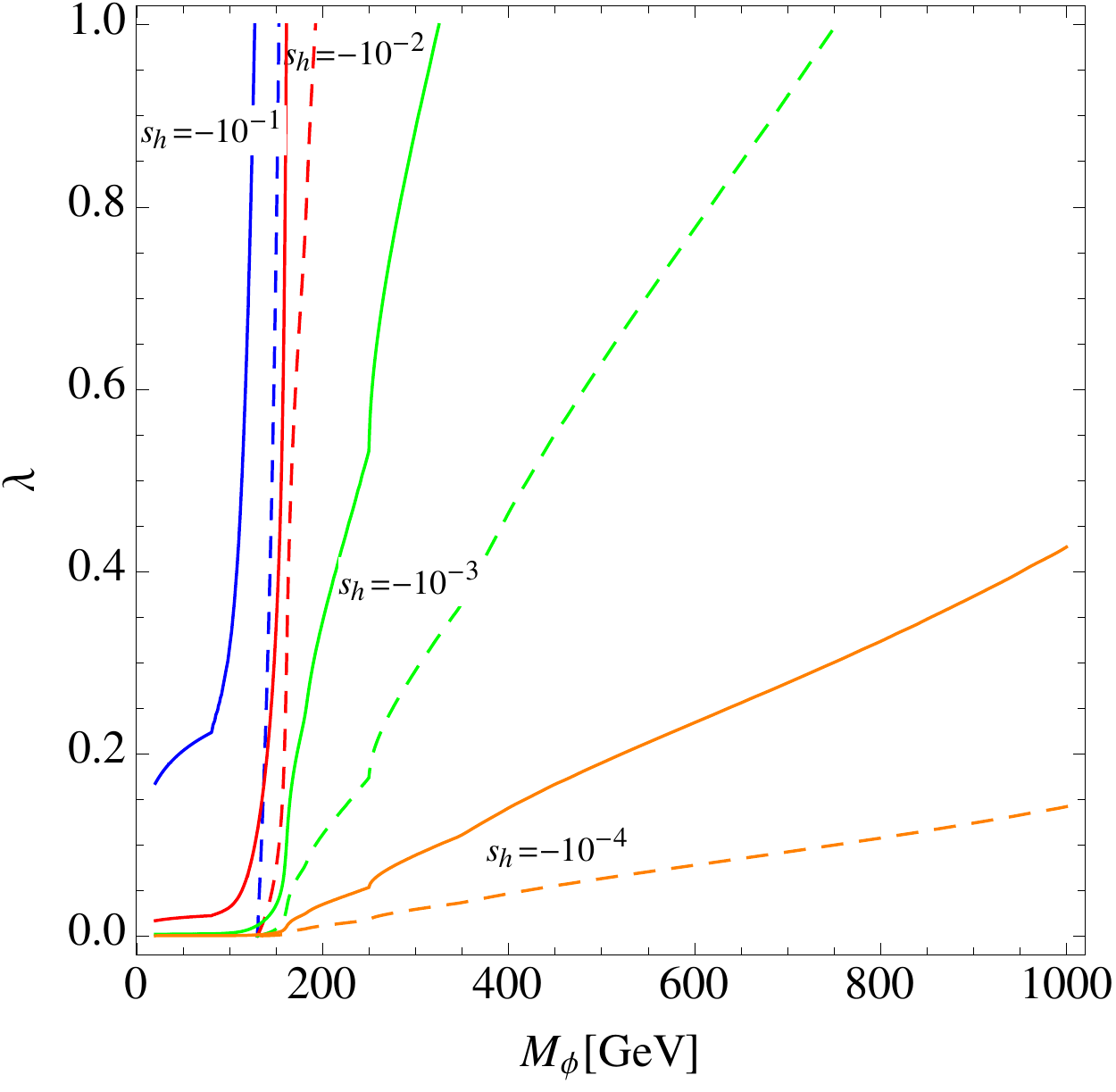}  
   \caption{Contours of $Br(\phi \rightarrow b\overline{b}$) = 0.9 (solid) and 0.5 (dashed), for $|s_h| = 10^{-1}$ (blue), $10^{-2}$ (red),   $10^{-3}$ (green),   and $10^{-4}$ (orange).  The left and right plots have $s_h > 0$ and $s_h<0$, respectively. 
   We take $M_B=1$ TeV and $w = M_\phi$, and we assume that $\phi \rightarrow Z' Z'$ is kinematically forbidden. 
   }
   \label{fig:phi_to_bbbar_BR}
\end{figure}

If the heavy $B$ quarks decay mainly to $\phi b$, the results of Section~\ref{sec:Results} will  apply when $\phi$ particles decay mainly to $b \overline{b}$.  Taking $M_B=1$ TeV and $w = M_\phi$, we show in Figure~\ref{fig:phi_to_bbbar_BR} the parameter space where $\phi \rightarrow b\overline{b}$ dominates.  When $\phi$ is sufficiently heavy to decay to $W$, $Z$, and $h$ pairs, the mixing in the scalar sector must be very small for $b \overline{b}$ to dominate over  these modes. We assume $M_{Z'} > M_\phi/2$ to make Figure~\ref{fig:phi_to_bbbar_BR},  but 
 $\phi \rightarrow Z'Z'$  can easily be the most important decay mode if it is kinematically accessible.

\subsection{Dark Matter}\label{sec:darkmatter}

In this work we focus mainly on the LHC phenomenology of the $B$ and the $Z'$.  
However, our model, over part of the parameter space, also provides a natural explanation for the excess of high energy gamma-rays seen coming from the proximity of the Galactic center, the so called Galactic Center Excess (GCE), or Gooperon \cite{Goodenough:2009gk,Hooper:2010mq,Boyarsky:2010dr,Hooper:2011ti,Linden:2012iv,Abazajian:2012pn,Hooper:2013rwa,Gordon:2013vta,Abazajian:2014fta,Daylan:2014rsa,Zhou:2014lva,Calore:2014xka,Agrawal:2014oha}.  The spectrum of the excess photons is well fit by a 30 -- 50 GeV DM particle annihilating directly to $b\bar{b}$, as well as by a 10 GeV DM particle annihilating to $\tau$'s.  It may also be fit by cascade annihilations of DM to light mediators which in turn decay to pairs of SM particles \cite{Boehm:2014bia,Ko:2014gha,Abdullah:2014lla,Martin:2014sxa}.  In particular, the spectrum of the GCE is better fit for annihilations of the form $\chi\chi\rightarrow \zp\zp\-\rightarrow (b\bar{b})(b\bar{b})$ than for direct annihilations to $b$'s if $M_\chi\sim 30\;\GeV + M_{\zp}/2$ \cite{Martin:2014sxa}, \eg  $(M_\chi,M_\zp) = (60, 50)\;\GeV$.

We introduce a pair of vector-like fermions, $\chi, \chi^c$, with charges $Q_\chi$ and $-Q_\chi$ under the $U(1)^\prime$ but no SM charge. 
Provided  $Q_\chi \neq 0$, these fermions are stable at the level of renormalizable interactions.  
Recall that we have normalized the $U(1)'$ gauge coupling $g'$ so that $\Phi$, $B$, and $B^c$ have charges  $+1$, $-1$, and $+1$.  If $Q_\chi$ is not an integer, an unbroken global, abelian symmetry guarantees the stability of  $\chi, \chi^c$.  Even if $\chi, \chi^c$ are not absolutely stable, they can easily be stable on cosmological time scales if any non-renormalizable operators that induce their decays are generated at the Planck scale or some other very high scale.  Provided $Q_\chi \neq \pm
1,\pm2$, there are no operators at dimensions five or six leading to $\chi$ decays.

For $M_\chi> M_{\zp}$ or $M_\chi > (M_{\zp}+M_\phi)/2$, the annihilation processes $\chi\overline{\chi}\rightarrow \zp\zp$ or $\chi\overline{\chi}\rightarrow \phi \zp$ are accessible.  This allows for a secluded DM scenario \cite{Pospelov:2007mp}, in which the couplings that determine the relic abundance are independent of those that determine the DM's coupling to the SM. 
Focussing on  $\chi\overline{\chi}\rightarrow \zp\zp$, the non-relativistic DM annihilation rate is
\be
\sigma v_{\chi\overline{\chi}\rightarrow \zp\zp} =
 \frac{g^{\prime4}Q_\chi^4}{16 \pi M_\chi^2} 
\left(
1-\frac{M_{\zp}^2}{M^2_\chi}
\right)^{3/2}
 \left( 
1-\frac{M_{\zp}^2}{2M^2_\chi}
\right)^{-2}.
\ee
For  masses that fit the GCE the correct relic abundance is achieved for $g^\prime Q_\chi\sim 0.2$.  
We have checked this and other results from this section using 
\texttt{micrOMEGAs} \cite{Belanger:2013oya}.
If $\chi \overline{\chi} \rightarrow \phi Z'$ is also a relevant annihilation channel, slightly smaller values of $g' Q_\chi$ work.

If  $M_\chi$ is too light to annihilate into final states involving $Z'$ and $\phi$, the correct relic abundance can still be achieved through $\chi\overline{\chi}\rightarrow b \overline{b}$, mediated by $s$-channel $Z'$ exchange.  
Neglecting mixing in the LH quark sector, the non-relativistic DM annihilation rate for this process is
\be
\sigma v_{\chi\overline{\chi}\rightarrow b \overline{b}} =
 \frac{3 s_R^4 g^{\prime4}Q_\chi^2}{2\pi}\frac{ M_\chi^2}{ \left( M_{\zp}^2 - 4 M_\chi^2\right)^2 } 
\left(
1-\frac{M_{b}^2}{M^2_\chi}
\right)^{1/2}
 \left( 
1-\frac{M_{b}^2}{4M^2_\chi}
\right)~.
\ee
With the help of  Eqn.~(\ref{eqn:fermion_sine_relation}), it is useful to rewrite this as
\be
\sigma v_{\chi\overline{\chi}\rightarrow b \overline{b}} =
 \frac{3 \lambda^4 Q_\chi^2}{8\pi}\frac{ M_\chi^2}{M_B^4}
 \left( 
1-\frac{4 M_{\chi}^2}{M^2_{\zp}}
\right)^{-2}
\left(
1-\frac{M_{b}^2}{M^2_\chi}
\right)^{1/2}
 \left( 
1-\frac{M_{b}^2}{4M^2_\chi}
\right).
\ee
Unlike the case where the relic abundance is set by $\chi\overline{\chi} \rightarrow Z' Z'/\phi Z'$, achieving the correct relic abundance through $\chi\overline{\chi}\rightarrow b \overline{b}$ requires that $M_B$ not be too large. The annihilation rate is resonantly enhanced for $M_{Z'}$ close to $2 M_\chi$, but the correct relic abundance can also be obtained far off resonance.  For example, taking  $M_B = 1$ TeV, $M_\chi = 40$ GeV (as preferred for the GCE), and $M_{Z'} =250$ GeV, we need $Q_\chi \lambda^2\simeq 4$.  Taking $\lambda=1$ and maximal mixing in the RH quark sector, we get  $g' = ( \lambda M_{Z'})/(\sqrt{2} s_R M_B) = 1/4$, and  the coupling of the $Z'$ to DM is not too large: $g' Q_\chi \simeq 1$.  

For $M_\chi< M_{\zp}/2$, the presence of DM coupled to the $\zp$ opens up an invisible decay mode with partial width
\be
\Gamma(\zp \rightarrow \chi \bar{\chi} ) = \frac{g^{\prime 2} Q_{\chi}^2}{12 \pi} \sqrt{1-4y_{\chi}} \left(1+2y_{\chi}\right)M_{\zp}~,
\ee
where $y_\chi = M_{\chi}^2/M_{\zp}^2$.  
The invisible width can easily dominate over the width into $b\overline{b}$, Eqn.~(\ref{eq:Zprimeqm1}).  In this case $B\overline{B}$ pair production at the LHC can lead to  $b\overline{b} + \met$  events  targeted by standard SUSY searches \cite{Aad:2013ija, Khachatryan:2015wza}.

Because the nucleus has no net $b$-charge, direct detection rates are highly suppressed in the absence of kinetic mixing.  Kinetic mixing leads to a spin-independent coupling of DM to the proton,  and to a cross section per nucleon 
\be
\sigma \approx \left(\frac{Z}{54}\right)^2\left(\frac{131}{A}\right)^2 \left(\frac{g^\prime Q_\chi}{0.2}\right)^2 \left(\frac{\kappa}{10^{-4}}\right)^2
\left(\frac{50\,\GeV}{M_{\zp}}\right)^4 \times 10^{-46} \mathrm{cm}^2~,
\ee
where we normalise to scattering off xenon. For $M_{\chi} \sim 50$ GeV, LUX  has probed down to $\sigma \simeq 8\times 10^{-46}$ cm$^2$ \cite{Akerib:2013tjd}. Parameters chosen to explain the GCE in the secluded DM scenario thus require $\kappa \lsim 3 \times 10^{-4}$ to evade direct detection.  Values  of $\kappa$ this small are not unreasonable, especially given that $g'$ can be small.
Taking $\kappa$ to be given by Equation~(\ref{eqn:kinmix}) with the log set to one, the constraint is satisfied for $g' \sim 1/20$, which requires $Q_\chi \sim 4$ for the relic abundance.  
The $\chi \overline{\chi} \rightarrow b\overline{b}$ explanation of the GCE is consistent with values of $M_{Z'}$ larger  than those preferred by the secluded DM explanation, meaning that LUX constraints can be satisfied with larger values of $\kappa$.

We have been assuming that $\chi$ and $\chi^c$ form a Dirac fermion of mass $M_\chi$, but it is possible that the mass eigenstates are Majorana fermions.  For example, if $Q_\chi = -1/2$, the interactions 
\be
{\mathcal L}   \supset  \lambda_\chi \chi \chi \Phi+\lambda_{\chi^c}  \chi^c \chi^c \Phi^* + h.c.
\ee
are allowed, leading to Majorana masses when $\Phi$ gets a vev.  If these Majorana masses are much smaller than the Dirac mass,  the relic density calculation does not change much, but the cross section for direct detection is dramatically reduced.  Larger values of $\lambda_\chi$ and/or $\lambda_{\chi^c}$ can change the phenomenology in various ways, {\em e.g.} scalar mixing can induce a Higgs-mediated contribution to the cross section for direct detection, final states involving $\phi$ can become more important for the relic abundance calculation, and $\phi$ can potentially decay invisibly to DM.

\section{Searching at the LHC}
\label{sec:searching}

Traditional searches for heavy vector-like $B$ quarks have focused on decays to SM bosons and quarks \cite{CMS:2013una,Aad:2015mba,atlasVLQ}.  
As we have seen, the presence of $Z'$ and $\phi$ (and $\chi$ if DM is included), can significantly alter the phenomenology.
 Which of the many possible search channels dominates depends upon the masses of 
 the new particles and upon the relative sizes of the various mixings, namely kinetic mixing, quark mixing, and scalar mixing.  

We will consider the situation where the dominant decays are $B \rightarrow Z' b$ followed by $Z' \rightarrow b \overline{b}$.  As discussed in Section~\ref{sec:themodel}, $B\rightarrow Z' b$ tends to dominate for $M_\phi > M_B >M_{Z'}$, unless $\langle\Phi\rangle$ is much larger than $M_B$ (see Figure~\ref{fig:Bratios}), while $Z' \rightarrow b \overline{b}$ dominates for $M_B > M_{Z'}$ and sufficiently small kinetic mixing (see Figure~\ref{fig:kinmix_vs_quarkmix}).  It will be possible to infer from our final results the effect of branching ratios smaller than one.   If $B$ decays to both $\zp b$ and $\phi b$ our analysis would find both resonances but at reduced significance, as long as both $\zp$ and $\phi$ decay to $b\overline{b}$. 

\begin{figure}[t] 
   \centering
   \includegraphics[width=0.5\textwidth]{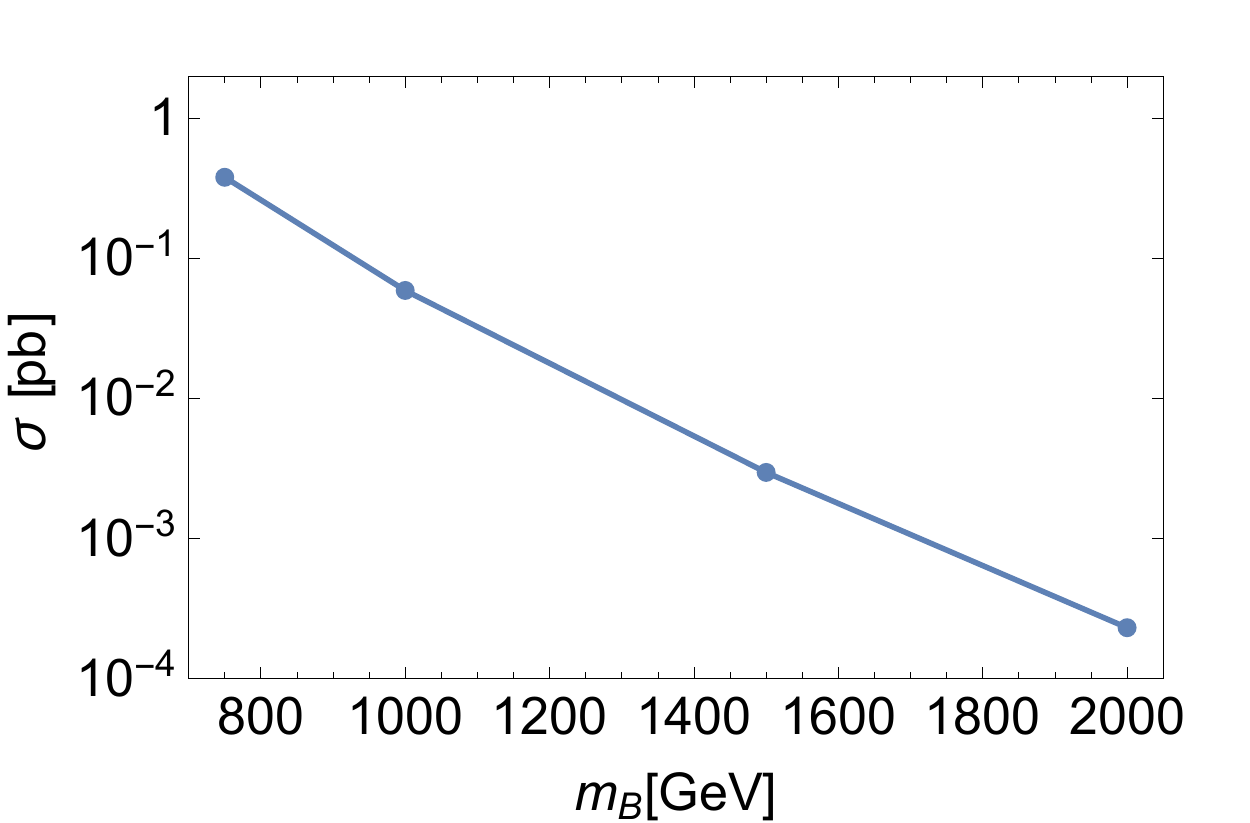} 
   \caption{$B\overline{B}$ production cross section at the LHC with $\sqrt{s}=14$ TeV.}
   \label{fig:BBdiag}
\end{figure}

The sizeable QCD production rate of $B\overline{B}$, shown in Figure~\ref{fig:BBdiag}, makes our primary channel of interest  $pp\rightarrow B\overline{B}\rightarrow (b\zp)(\bar{b}\zp)\rightarrow b(b\bar{b}) \bar{b}(b\bar{b})$, which is not presently being searched for.  Before describing in detail the search strategy we advocate, we briefly discuss other interesting channels that
are worthy of investigation.

Although their couplings are  suppressed by the quark mixing angle, the $\zp$ and $\phi$ can be singly produced in association with $b$ quarks, which may be forward boosted. If these  states decay to $b\overline b$, their existence is probed by LHC searches for $b\bar{b}$ resonances produced in association with $b$ quarks \cite{Khachatryan:2015tra}.  

With kinetic mixing the $\zp$ will have a di-leptonic branching ratio, but unless $\kappa$ is sufficiently large the usual $\zp$ bounds are weakened by the necessity of producing it in association with $b$ quarks.  The dilepton resonance can also show up in decays of the $B$, in which case the final state would be a pair of dileptonic resonances and two $b$ quarks, which can be paired up into two $b\ell\ell$ resonances.

If $\phi$ is sufficiently heavy it can decay to $Z'Z'$.  Or, if $\phi\rightarrow Z'Z'$ is kinematically forbidden 
but the scalar mixing is sufficiently large, then $\phi$ can decay to $hh$, $WW$, and $ZZ$  if it is heavy enough.  
When $B \rightarrow \phi b$ dominates, $B \overline{B}$ production can therefore lead to events with as many as ten $b$ quarks, with various sub-resonances among the $b$-jets.  Finally, if we incorporate DM into the theory the $\zp$ and/or the $\phi$ might decay invisibly, leading to events with $b$-jets and MET.

Returning to our channel of primary focus, $B\overline{B}\rightarrow (b\zp)(\bar{b}\zp)\rightarrow b(b\bar{b}) \bar{b}(b\bar{b})$, the results of  Section~\ref{sec:Results}  are based on simulations of 45 separate parameter points covering
 a broad range of $B$ and $\zp$ masses.  Before presenting those results, we describe our analysis technique.
 To aid in the discussion, we adopt three representative benchmark points.

\subsubsection*{Benchmark 1 ($M_B = 1$ TeV, $M_{Z'} = 50$ GeV)} 

This light $\zp$ benchmark is motivated by the secluded DM explanation of the GCE if, as discussed in Section~\ref{sec:darkmatter}, the DM mass is around 60 GeV.  Larger values of $M_{Z'}$ are consistent with the GCE if the dark matter annihilates directly to $b\overline{b}$.    This benchmark requires jet-substructure techniques because the large mass difference between $B$ and $\zp$ means that the $b\bar{b}$ from the $\zp$ decay  will typically form a single massive jet.  

It is not difficult to find parameters consistent with  $M_B = 1$ TeV,  $M_{\zp} = 50$ GeV,  and  $Br(B \rightarrow Z' b) \simeq Br(Z' \rightarrow b \overline{b}) \simeq 1$.  For example, start with  $g' = 1/20$, corresponding to $\langle \Phi \rangle=M_B/\sqrt{2}$ and $s_R = \lambda/\sqrt{2}$.  For this value of $g'$, $B\rightarrow \phi b$ is forbidden if the $\Phi$ quartic coupling satisfies $\lambda_\Phi >1/2$ (here we neglect scalar mixing), in which case Figure~\ref{fig:Bratios} shows the $B$ decays dominantly to $Z' b$ (unless $s_R \simeq 1$).  Figure~\ref{fig:kinmix_vs_quarkmix} shows that for $\kappa \lsim 10^{-2} \lambda^2$, $Z'$ will mainly decay to $b\overline{b}$.  
If we incorporate Dirac fermion dark matter with $M_\chi = 60$ GeV, the relic abundance  requires  $g^\prime Q_\chi\approx 0.2$ in the secluded DM scenario, or $Q_\chi \approx 4$.  Then we need $\kappa \lsim 3\times 10^{-4}$ to satisfy direct detection constraints.

\subsubsection*{Benchmark 2 ($M_B = 1.5$ TeV, $M_{Z'} = 750$ GeV)} 

This ``medium mass" point can be discovered with high significance after 300 fb$^{-1}$ of data, even with sizable systematic uncertainties, and will have hints after 30 fb$^{-1}$ (see Figure~\ref{fig:significances}).  
An example set of model parameters for this point starts with  $\langle\Phi\rangle = 1500$ GeV (corresponding to $g' = 0.35$ and $s_R = \lambda$).  With this choice of parameters, $M_\phi> M_B$ is realized for $\lambda_\phi\gtap 1/4$, in which case $B\rightarrow Z'b$ typically dominates.  For $Z'\rightarrow b\overline{b}$ to dominate only requires
$\kappa \lsim 0.16\lambda^2$.

\subsubsection*{Benchmark 3 ($M_B = 2$ TeV, $M_{Z'} = 1.5$ TeV)} 

Because of its  small production cross section, this ``high mass" point may require as much as 3000 fb$^{-1}$ to be discovered.   For an example set of parameters we can again start with $\langle\Phi\rangle = 1500$ GeV (corresponding to $g' = 1/\sqrt{2}$ and $s_R= 3\lambda/4$).   To have $M_\phi> M_B$ we need $\lambda_\phi\gtap 4/9$, and for $Z' \rightarrow b\overline{b}$ to dominate we need $\kappa\ltap 0.19 \lambda^2$.

\subsection{Simulation}  We implement the model in \texttt{Feynrules} \cite{Alloul:2013bka}.  Our signal simulations use  \texttt{MadGraph5\char`_aMC$@$NLO} \cite{Alwall:2014hca} for parton-level event generation,  \texttt{PYTHIA\_8.2} \cite{Sjostrand:2007gs} for showering and hadronization, and  \texttt{Delphes3} \cite{deFavereau:2013fsa} for detector simulation. The dominant background comes from QCD multijet production, followed by $t \overline{t}$ production.   We simulate these background processes with \texttt{PYTHIA\_8.2} and \texttt{Delphes3}.  Jets are clustered with FastJet \cite{Cacciari:2011ma} using the anti-kt algorithm \cite{Cacciari:2008gp} with $R=0.5$. For Delphes settings we use the default ``CMS'' parameter card that comes with the distribution.    This card sets the $b$-tagging efficiencies for the high-$p_T$ jets that will be important for our analysis at approximately 0.5 ($| \eta | \le  1.2$) and 0.4 ($1.2 < | \eta | \le  2.5$) for $b$-jets, 0.2 ($| \eta | \le  1.2$) and 0.1 ($1.2 < | \eta | \le  2.5$) for $c$-jets, and $10^{-3}$ for light jets.  

 We use  \texttt{Hathor} \cite{Aliev:2010zk} to calculate vector-quark production cross sections at NNLO \cite{Czakon:2013goa} with MSTW2008 NNLO parton distribution functions \cite{Martin:2009iq}.  For the $t \overline{t}$ production cross section we take $\sigma = 954$ pb,  based on Ref.~\cite{Czakon:2013goa}.   For the QCD background we adopt the LO cross section reported by Pythia.  Pythia8 with default settings has been found to give reasonable agreement, at a level better than $\sim$ 50\%,  with 7 TeV LHC data on multijet production  \cite{Aad:2011tqa,Karneyeu:2013aha}.    The difficulty in modeling the QCD background requires that it be estimated from data in an actual analysis.  We discuss one approach to this estimation in Section~\ref{sec:Results}.

To reduce the statistical uncertainty associated with our QCD simulations, we bias the event generation to favor high-$p_T$ events and record the event weights. We estimate the statistical uncertainties of our QCD Monte Carlo sample as 
\be
\frac{\sqrt{\sum_i w_i^2}}{\sum_i w_i},
\ee
where the $w_i$ are the individual event weights.  This uncertainty is less than 10\% for most of the signal windows we use to obtain the results of Section~\ref{sec:Results}.

\subsection{Analysis}

\begin{figure}[t] 
   \centering
   \includegraphics[width=0.45\textwidth]{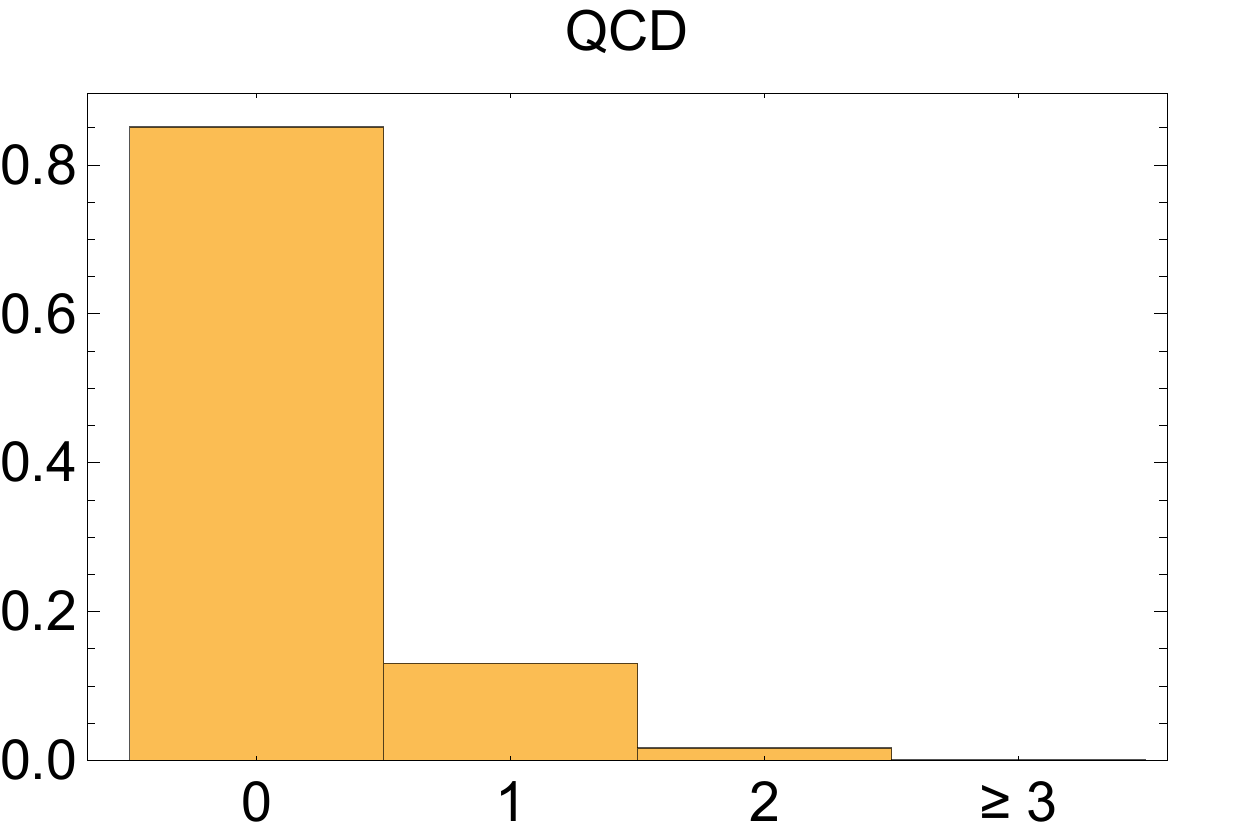} 
    \includegraphics[width=0.45\textwidth]{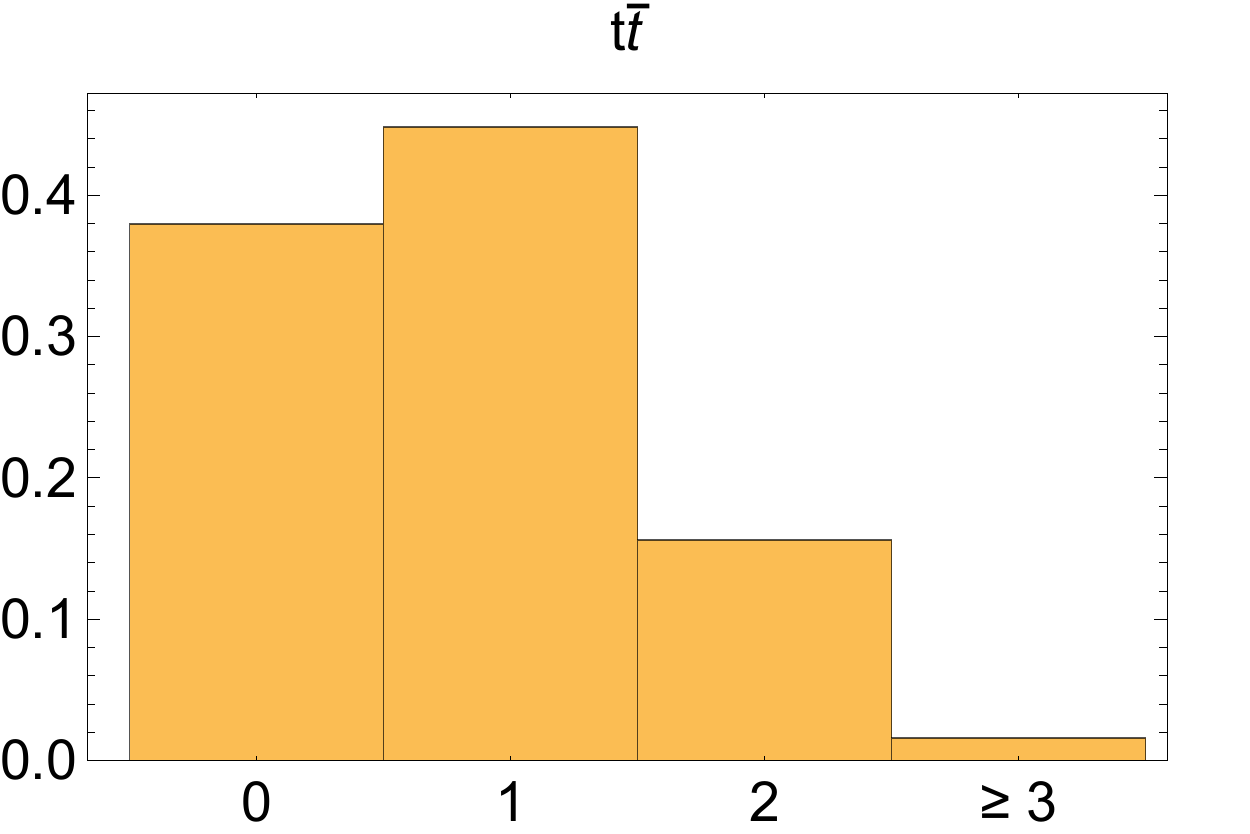}  
   \caption{Probabilities to have 0, 1, 2, and 3 or more $b$-jets, 
  among background events with at least four jets and $H_T > 500$ GeV.  For QCD events the probability to have at least 3 $b$-jets is $1.2 \times 10^{-3}$. }
   \label{fig:btagrates-background}
\end{figure}

\begin{figure}[t] 
   \centering
   \includegraphics[width=0.32\textwidth]{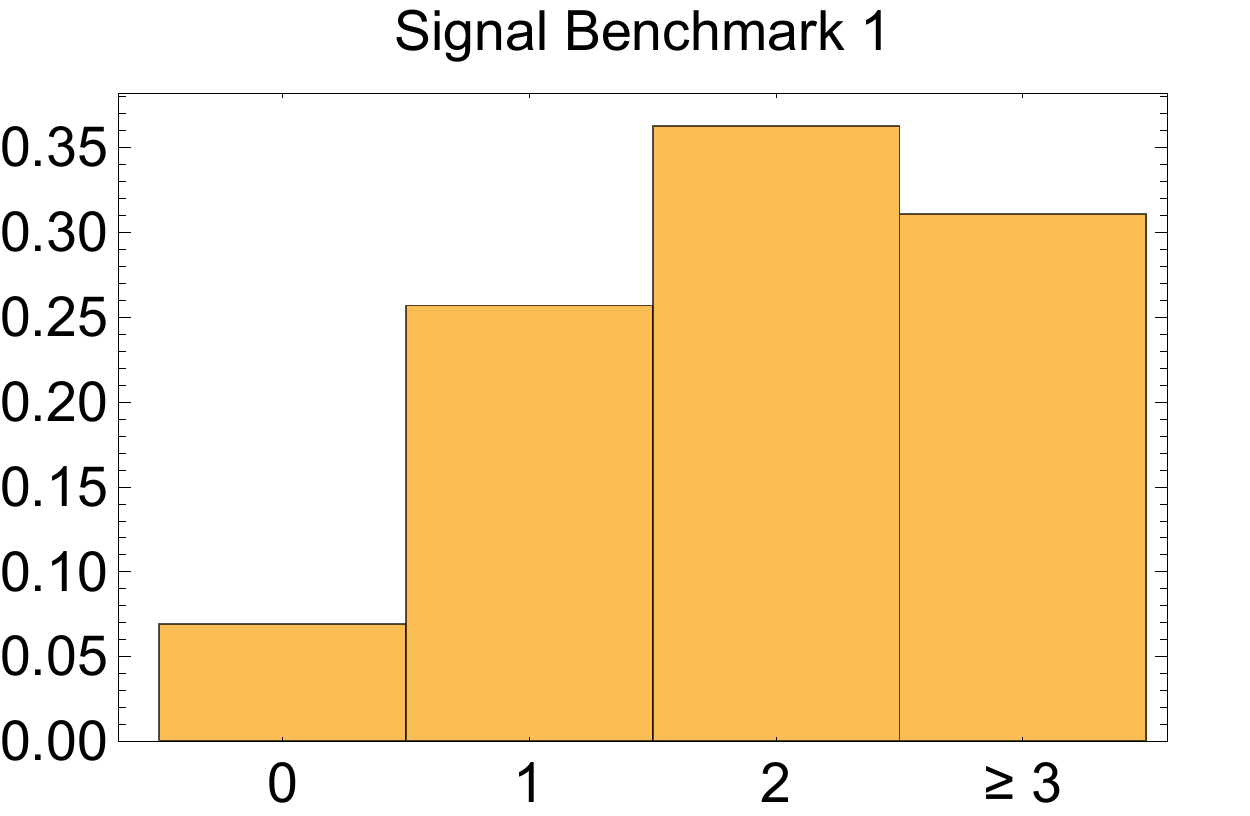} 
    \includegraphics[width=0.32\textwidth]{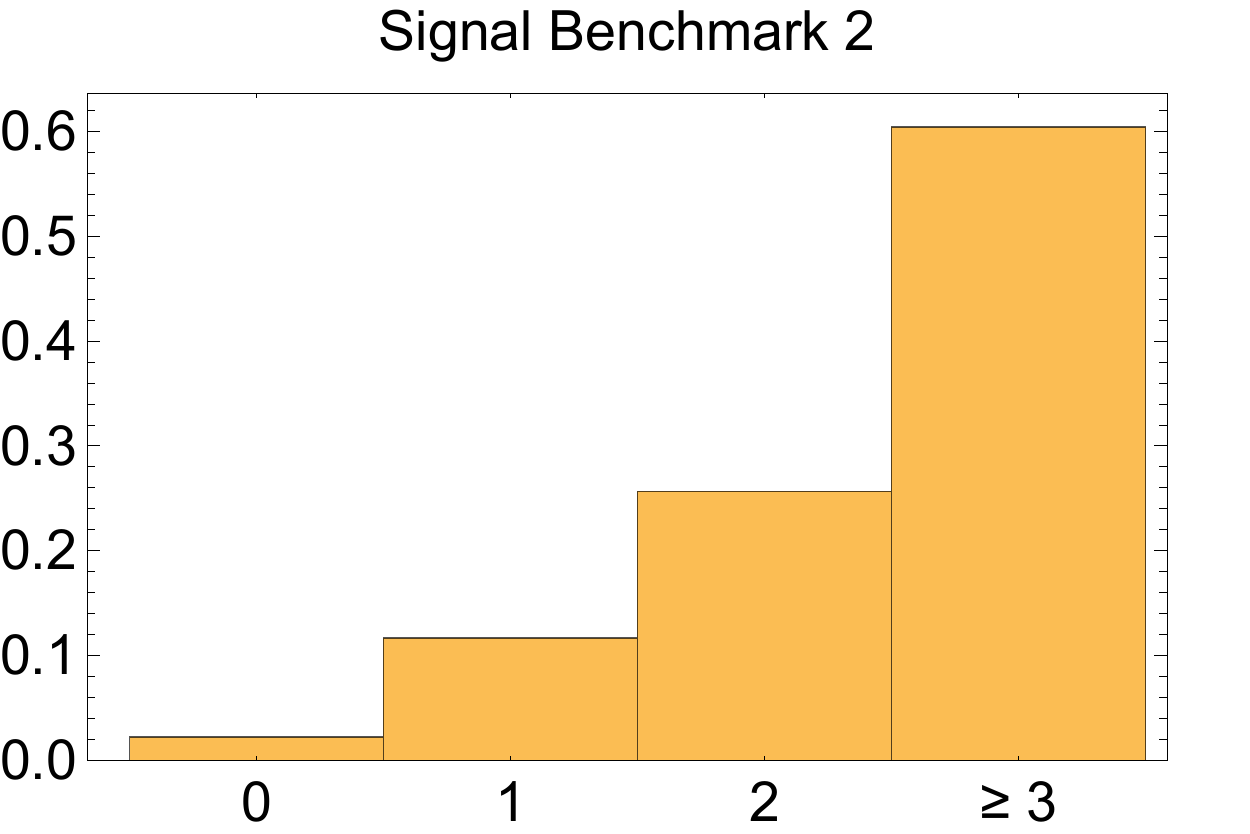}  
    \includegraphics[width=0.32\textwidth]{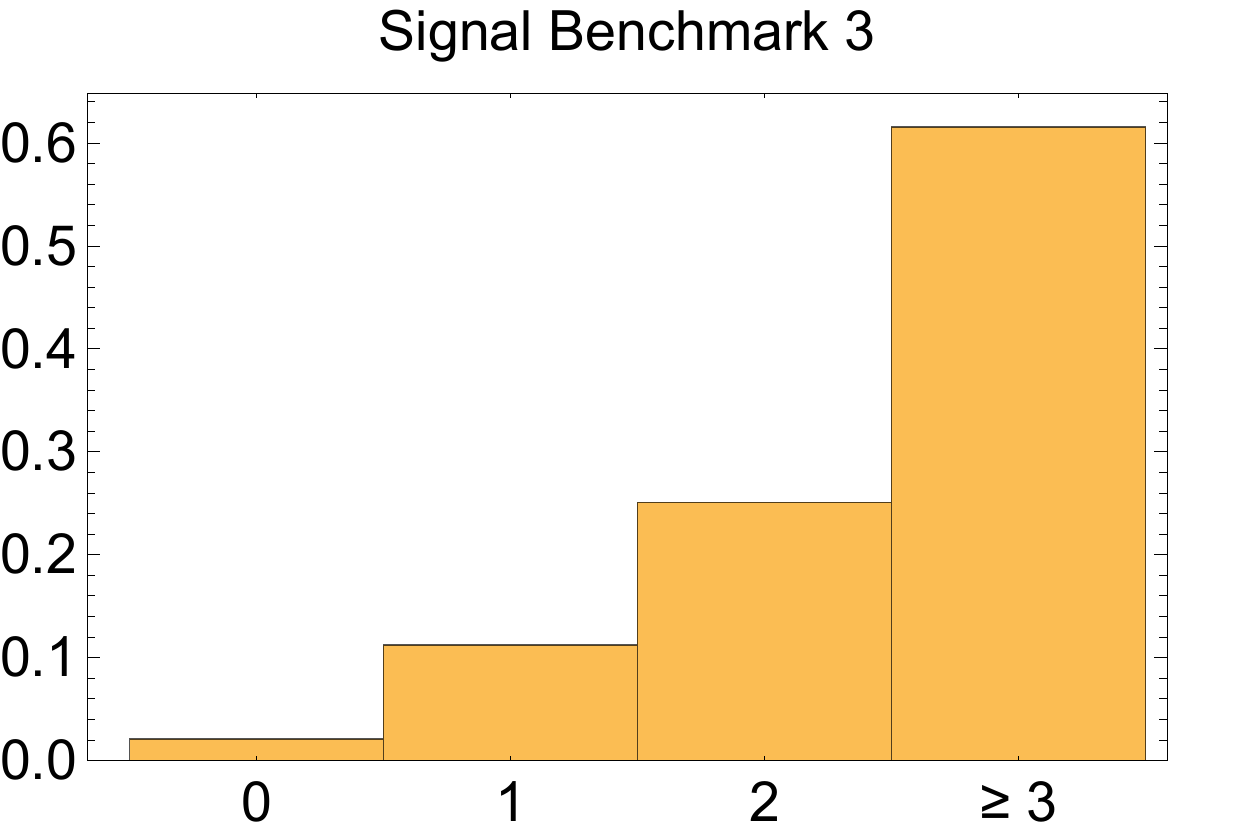}  
   \caption{Probabilities to have 0, 1, 2, and 3 or more $b$-jets, 
  among background events with at least four jets  and $H_T > 500$ GeV. The signal benchmarks are described in the text.}
   \label{fig:btagrates-signal}
\end{figure}

Only jets with $p_T > 100$ GeV and $|\eta| < 2.5$ are considered in our analysis.  In the discussion that follows, ``jet'' refers to an object satisfying these criteria, and we calculate the scalar sum of jet $p_T$'s, $H_T$,  using only these jets.  To be selected, an event must have at  least four jets  ($n_j \ge 4$),  three or more of which must be $b$-tagged ($n_b \ge 3$).    The probabilities to have various $n_b$, among events with $n_j\ge 4$ and $H_T > 500$ GeV, are shown in Figures \ref{fig:btagrates-background} and \ref{fig:btagrates-signal} for the backgrounds and for the three benchmark points introduced above. 

Figure \ref{fig:btagrates-signal} shows a lower probability to satisfy the $n_b \ge 3$  requirement for Benchmark 1, because  $B$ decays produce highly boosted $Z'$ particles for this parameter point, leading to $Z'$ decays that typically produce a single jet.   A more sophisticated analysis might attempt to keep track of the number of $b$-tags associated with individual jets.  Figures \ref{fig:btagrates-background} and \ref{fig:btagrates-signal} also suggest that it may be advantageous to require more than three $b$-jets, especially if one adopts a looser $b$-tag algorithm with a higher efficiency than we assume.   For examples of how requiring a high number of b-tags ($\ge 5$) may  be able to reduce backgrounds and allow discovery of certain signals, see Ref.~\cite{Evans:2014gfa}.  We present results for an analysis based on $n_b \ge 3$ to be conservative, and we will see that with this analysis there is discovery potential for $M_B = 2$ TeV at the HL-LHC.  

For each selected event we apply three separate reconstruction strategies.  These strategies differ in how many of the jets in the event are used in the reconstruction and in how $Z'$ candidates are identified.  Once $Z'$ candidates are found the identification of $B$ candidates proceeds identically for all three approaches.

The {\em four-jet reconstruction} uses only the four hardest jets in the event.  Among these four, two jets are identified as $Z'$ candidates if their jet masses match to within 10\% and both jets have $\tau_2/\tau_1 < 0.5$, where $\tau_N$ is the $N$-subjettiness variable defined in Ref.~\cite{Thaler:2010tr} .   This approach is effective for $M_B \gg M_{Z'}$, in which case the $Z'$ particles are produced with a large boost.   The {\em six-jet reconstruction} uses the six hardest jets in the event.  Among these six jets, two dijet pairs (comprising a total of four jets) are identified as $Z'$ candidates if the dijet masses match to within 10\%.  The {\em five-jet reconstruction} uses the hardest five jets and takes a composite approach.  Among the hardest five jets, a single jet and a dijet pair are identified as $Z'$ candidates if their masses match to within 10\% and the single jet has $\tau_2/\tau_1 < 0.5$.  

Regardless of which reconstruction method is applied to a particular event, there remain two available jets after two $Z'$ candidates are identified.  These  jets are paired with the $Z'$ candidates in both possible ways.  For each pairing, if the  jet-$Z'$ invariant masses are within 10\% of each other, then the  jet-$Z'$ systems are identified as $B$ candidates, and the two $(M_{Z'},M_B)$ pairs are recorded.  If $Z'$ candidates cannot be used to find $B$ candidates, then the $Z'$ candidates are discarded along with their associated masses.  

A single event may yield numerous  $(M_{Z'},M_B)$ pairs, produced by any and all of the three reconstruction methods.  Once we establish a range of $M_{Z'}$ and $M_B$ values as a useful window for a particular signal parameter point, we count an event as being in the window once and only once if any of its $(M_{Z'},M_B)$ pairs falls in that window.  This single counting allows for a more straightforward statistical interpretation of results.

\begin{figure}[t] 
   \centering
   \includegraphics[width=0.32\textwidth]{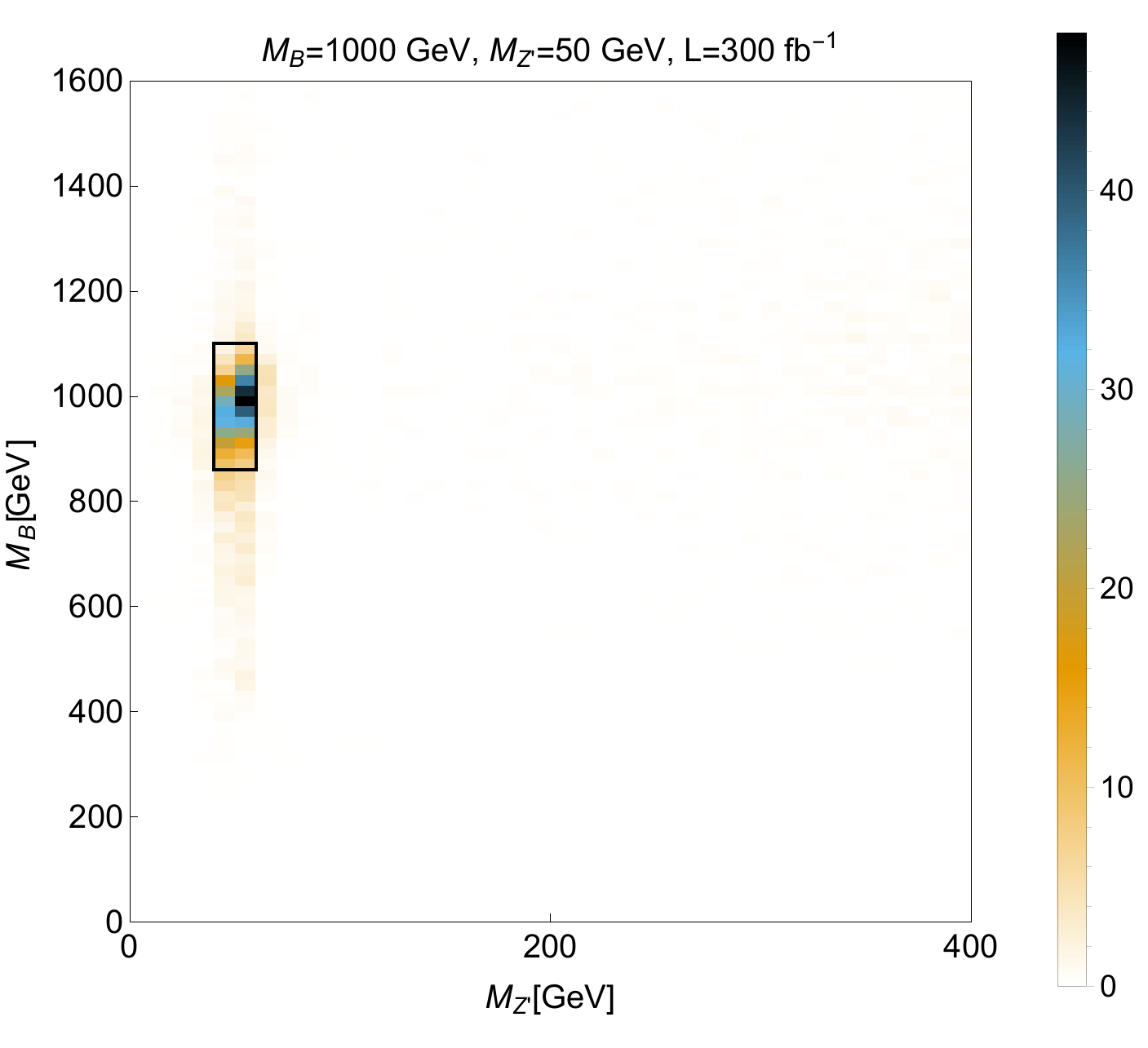} 
   \includegraphics[width=0.32\textwidth]{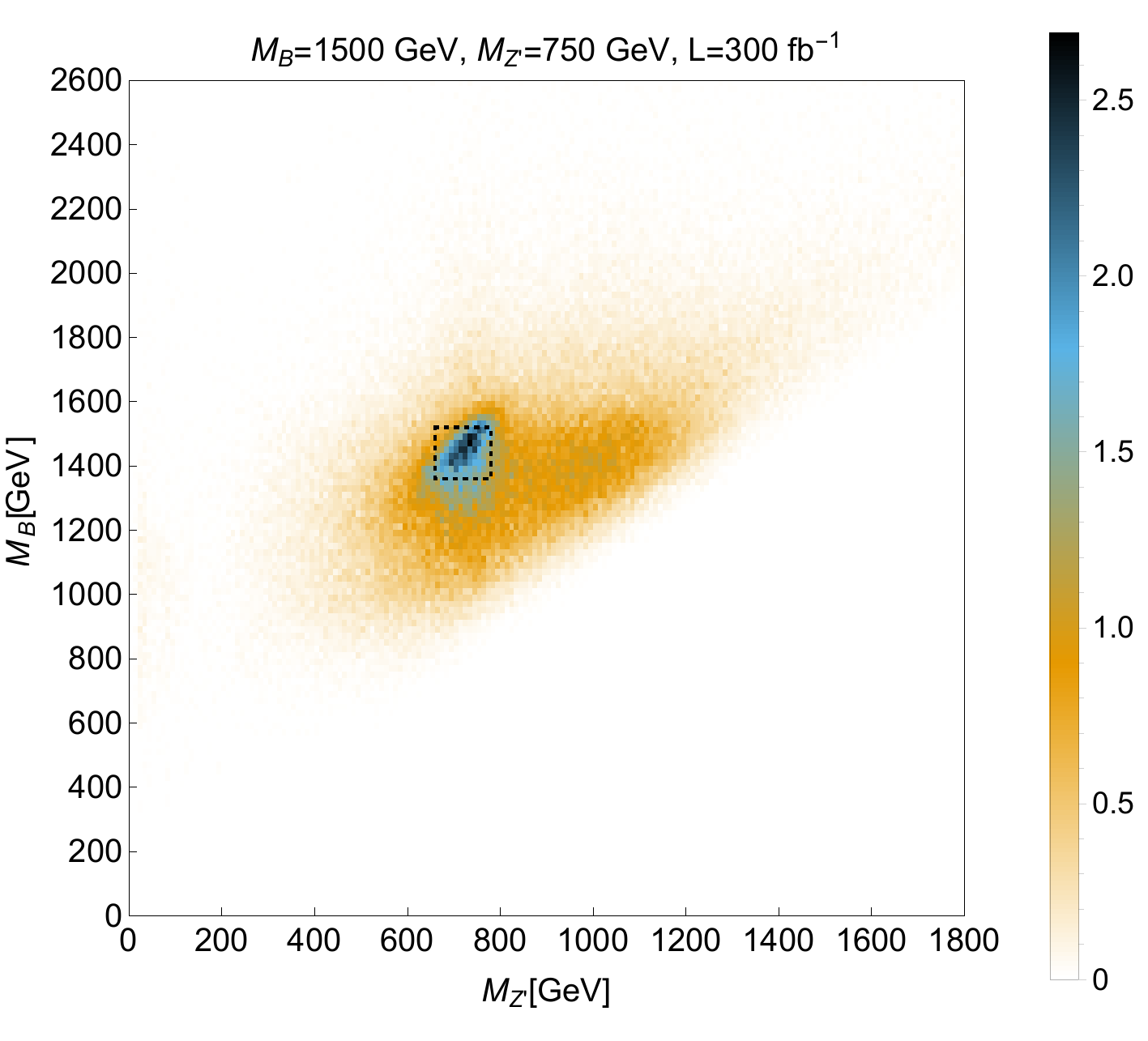}  
   \includegraphics[width=0.32\textwidth]{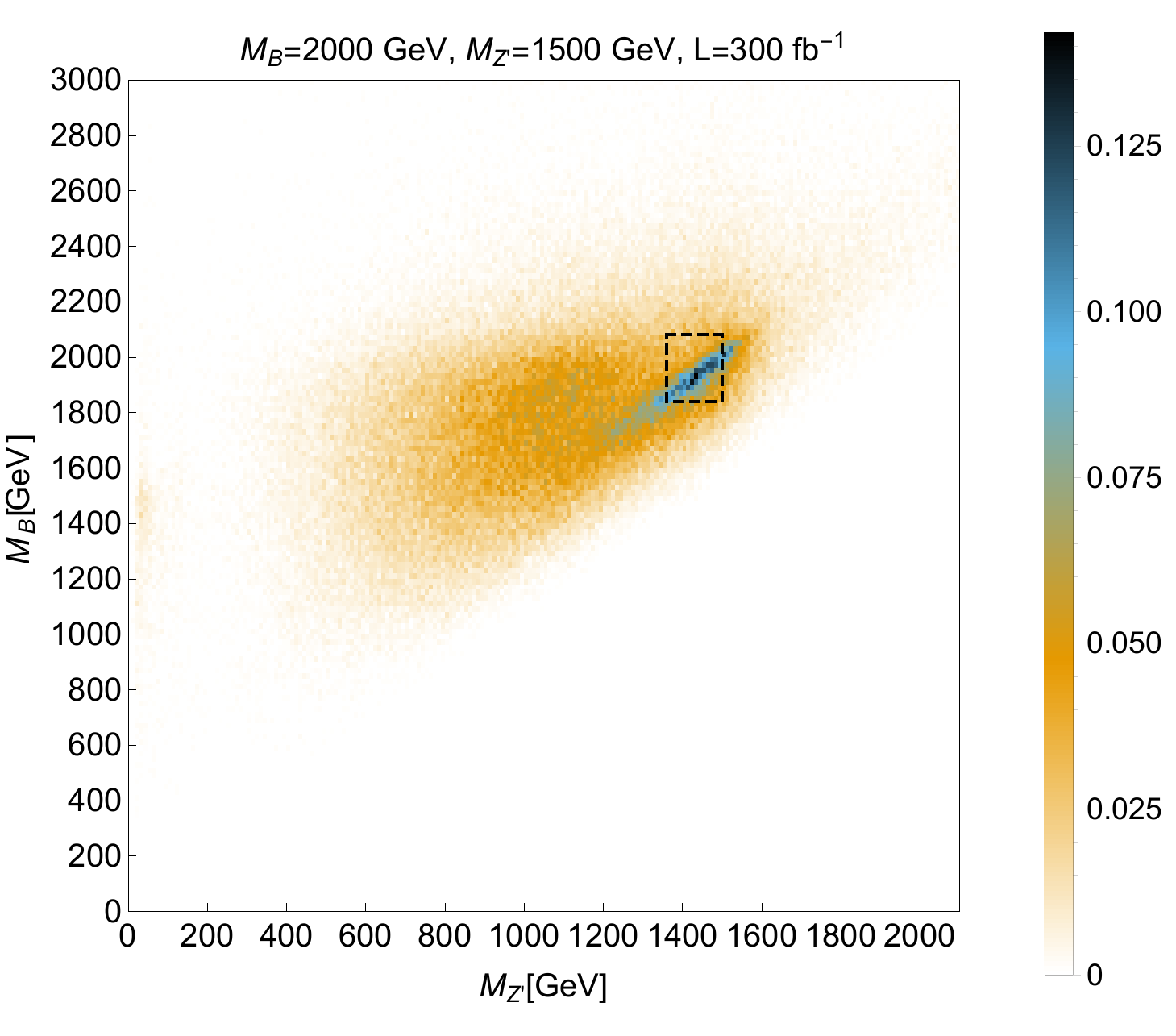}  
   \caption{Distribution of signal events satisfying $n_j \ge 4$, $n_b \ge 3$, and $H_T> 500$ GeV, for the three benchmark signals.  
   The rectangular selection windows are those of Table~\ref{tab:signalwindows}, optimized for  $\delta = 10\%$.}
   \label{fig:sigdistributions}
\end{figure}

\begin{figure}[htbp] 
   \centering
    \includegraphics[width=0.45\textwidth]{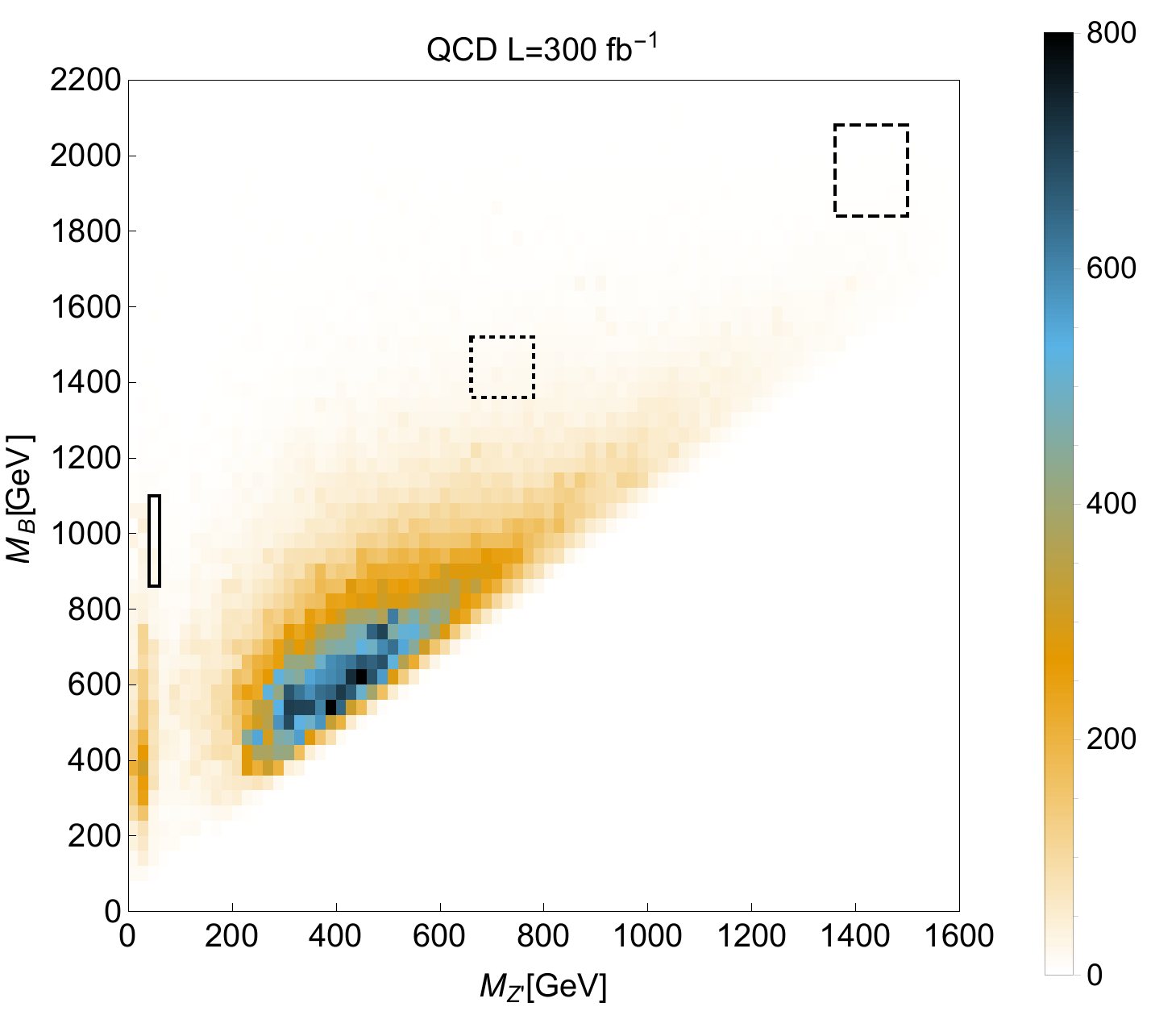} 
        \includegraphics[width=0.45\textwidth]{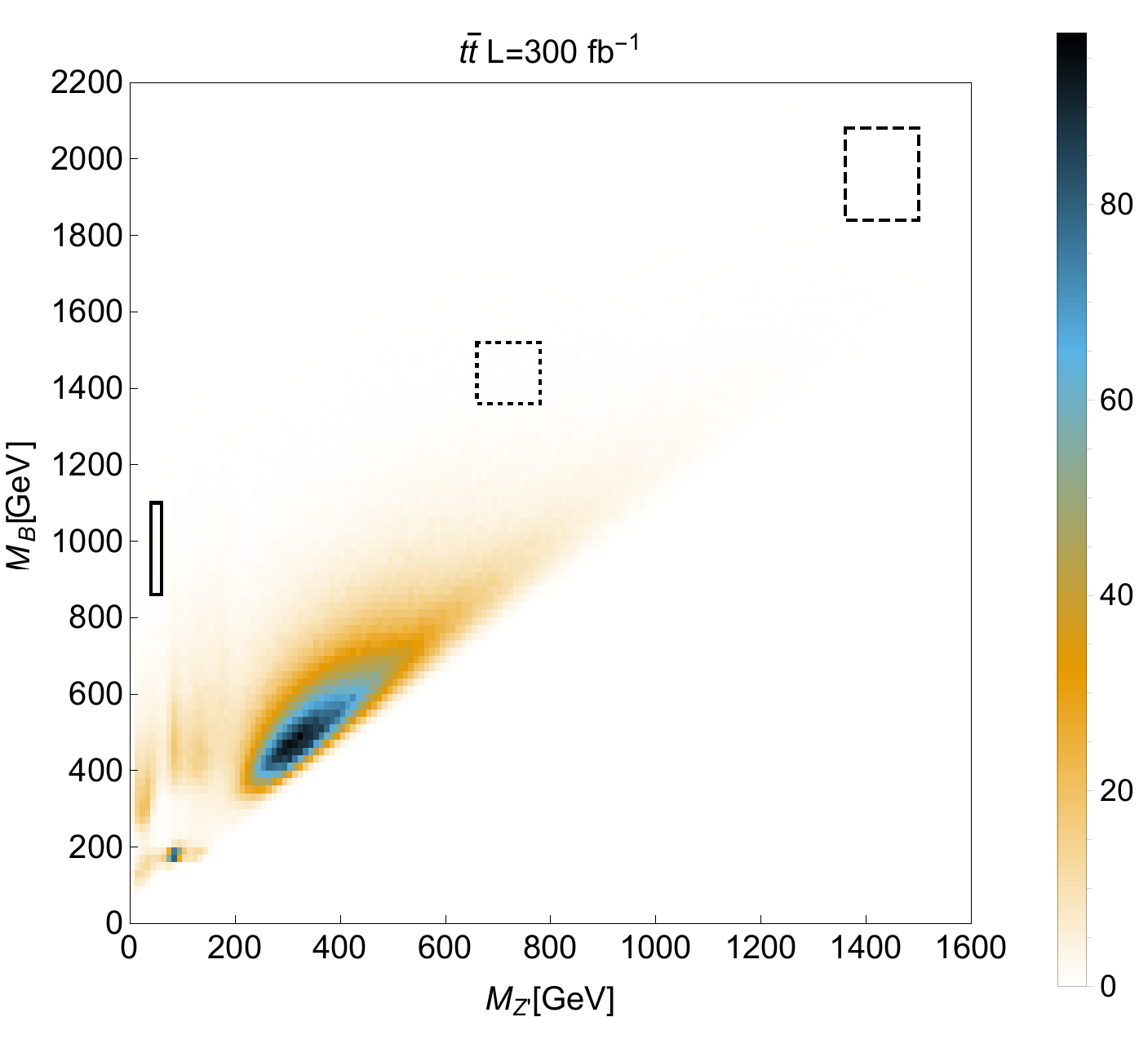}  
   \caption{
   Distribution of
   QCD (\emph{left}) 
   and $t\bar{t}$ (\emph{right}) events  satisfying $n_j \ge 4$, $n_b \ge 3$, and $H_T> 500$ GeV.  The pixels are 20 GeV $\times$ 40 GeV (\emph{left})  and 10 GeV $\times$ 20 
  GeV (\emph{right}) in size.  The three signal selection windows from Figure~\ref{fig:sigdistributions} are also shown.
   }
   \label{fig:backgrounddistributions}
\end{figure}

Figure~\ref{fig:sigdistributions} shows the distribution of signal events in the $M_{Z'}-M_B$ plane for our three benchmarks.  To make these plots we divide the $M_{Z'}-M_B$ plane  into 10 GeV $\times$ 20 GeV pixels.  A given event can count at most once in a given pixel but is allowed to be counted in multiple pixels.  
Similarly, Figure~\ref{fig:backgrounddistributions} shows the distribution of QCD and $t {\overline t}$ events in the $M_{Z'}-M_B$ plane.  In the $t {\overline t}$ plot, we see a concentration of events near $(M_W, M_t)$ due to successful reconstruction of the $W$ and top resonances.  We get much larger counts in a bulk region whose position is set by the $H_T$, jet $p_T$, and jet multiplicity requirements.  These are combinatorially favored ``mispairings'' produced by the six-jet reconstruction.  
Mispairings also produce additional concentrations at $M_B \gg M_t$,  with  $M_{Z'} \simeq M_W$ or with $M_{Z'}$ between $M_W$ and $M_t$.  Finally, the counts at very small values of reconstructed $M_{Z'}$ arise from the four-jet reconstruction, where individual jets with similar small jet masses can constitute a pair of $Z'$ candidates.

The $Z'/B$ search should be carried out in a way that covers as much of the $M_{Z'}-M_B$ plane as possible.  In Section~\ref{sec:Results} we present the results of the following strategy: for a given $M_{Z'}, M_B$ signal point to be tested, we impose a cut $H_T > 1.5 M_B$ and construct an appropriate window in the $M_{Z'}-M_B$ plane.  We set the boundaries of the window by a loose optimization of the quantity
\be
\frac{S}{\sqrt{S+B+(\delta B)^2}},
\ee
where $S$ and $B$ are the expected numbers of signal and background events in the window, and where $\delta$ represents the systematic uncertainty associated with  the background in the window.   Once a window is chosen, we quantify the expected significance of the signal using
\be
\frac{S}{\sqrt{B+(\delta B)^2}}
\ee
for $B\ge50$.  For smaller $B$, we take the background-model probability of observing $n$ counts as
\be
P(n) = \int \!d\lambda\; f(\lambda | B, \delta B) \;g(n|\lambda),
\ee
where $g(n|\lambda)$ is the Poisson distribution with mean $\lambda$ and $f(x | \mu, \sigma)$ is the normal distribution with mean $\mu$ and standard deviation $\sigma$, and we quantify the significance as
\be
\sqrt{2}\; \text{erf}^{-1} \left(1-2\sum_{n\ge S+B} P(n) \right).
\ee
We will present results for $\delta = 0$ and $\delta = 10\%$.   In the following section, we argue that estimating background from data at a 10\% level  or better is a realistic goal for this analysis.

\section{Results}

\label{sec:Results}

\begin{table}
\begin{tabular}[t]{||l|c|c|c||}
\hline
\hline
& Benchmark 1 & Benchmark 2 & Benchmark 3 \\
& [50 GeV, 1000 GeV] &  [750 GeV, 1500 GeV] & [1500 GeV, 2000 GeV] \\
\hline
Bottom-left corner & (30, 840); (40, 860) & (640, 1280); (660, 1360) & (1360, 1840); (1360, 1840) \\
Top-right corner & (70, 1120); (60, 1100)  & (780, 1560); (780, 1520) & (1500, 2080); (1500, 2080)\\
\hline
\hline
\end{tabular}
\caption{Benchmark signal windows optimized for $\mathcal{L}=300$ fb$^{-1}$, with all units in GeV.  The first entry is for 0\% systematics ($\delta = 0$), and the second entry is for 10\% systematics ($\delta = 10\%$).}
\label{tab:signalwindows}
\end{table}

We have studied the discovery prospects for 45 signal parameter points in all.  Tables~\ref{tab:signalwindows} and 
and \ref{tab:benchmark_results} provide detailed results for our three benchmark points.  Table~\ref{tab:signalwindows} gives the $M_{Z'}-M_B$ selection windows used for each benchmark, optimized for 
an integrated luminosity of $\mathcal{L}=300$ fb$^{-1}$, and for either $\delta = 0$ and $\delta = 10\%$.  The windows for $\delta = 10\%$ are shown in Figures~\ref{fig:sigdistributions} and \ref{fig:backgrounddistributions}.  Table~\ref{tab:benchmark_results} shows the numbers of events that pass the various cuts in our analysis, for background and for the three signal benchmarks.  

In an actual experimental analysis it will be important to estimate the QCD background from data.
The background in a given window can be estimated using events with fewer $b$-tagged jets.  For the  $\delta = 10\%$ selection windows of Table~\ref{tab:signalwindows},  Table \ref{tab:background_estimate} compares the numbers of background events that pass the full analysis with the estimate
\be
\sum_{n_j \ge 4} 
\left( \frac{\text{\#  with }n_j \text{ jets and } n_b \ge 3}{\text{\# with }n_j\text{ jets and }n_b  = 0} \right) \times 
\left( 
{\text{\#  in  window, with } n_j \text{ jets and }n_b  = 0} 
\right)~.
\ee
In the first factor, the events must pass the $H_T$ cut (which differs for the different benchmarks, as the $H_T$ cut is set to be $H_T>1.5 M_B$), but the events are not required to yield $Z'$ or $B$ candidates.  In the second factor, the events must pass the full analysis, with at least one pair of $Z'$ and $B$ candidates  with masses in the window, except that the usual requirement $n_b \ge 3$ is replaced with $n_b=0$.   

Instead of using $n_b=0$ events for the estimate, one could instead use $n_b<3$,  $n_b =1$, or $n_b=2$ events, which might be more accurate.   However, Table~\ref{tab:background_estimate} shows that using $n_b=0$ events works rather well for the benchmark windows,  and the signal contamination of the background in the $n_b=0$ samples is less than 1\% for all three windows.  

For most of the signal points we investigated, the accuracy of the estimate using $n_b=0$ events is comparable to the level  of agreement  shown in Table~\ref{tab:background_estimate}.  Exceptions include several of the points with $M_{Z'} = 100$ GeV, where the $t {\overline t}$ background makes up a larger component of the background then for other points, due to the presence of $W$'s.   However, these points are heavily signal-dominated, \ie\ they have a large $S/\sqrt{B}$.  If the background estimation is not quite as good as we assume, the discovery potential changes very little. Furthermore, other handles for estimating the background will be at experimentalists' disposal, including events with reconstructed $M_{Z'}$ and/or $M_B$ values outside the window, or perhaps events for which the mass-matching that identifies $Z'$ and/or $B$ candidates fails at 10\% but satisfies some less stringent requirement.

\begin{table}
\begin{tabular}[t]{||l|c|c|c|c|c||}
\hline
\hline
Cut & QCD & $t\bar{t}$ & Benchmark 1 & Benchmark 2 & Benchmark 3\\
\hline
4 jets, $H_T > 500\ \GeV$ & $8.9\times 10^8$  & $6.9\times 10^6$ & 14900 & 888 & 69.2 \\ 
$H_T>1500\ \GeV, n_b\ge 3$ & 47200 & 5400 & 3740 & 491 & 39.6\\
$H_T>2250\ \GeV, n_b\ge 3$ & 5550  & 643 & 1160 & 412  & 38.8\\
$H_T>3000\ \GeV, n_b\ge 3$ & 834 & 98.4 & 203 & 143  & 31.6\\
Mass pair within 10\% & $\begin{pmatrix}6850\\ 1030 \\ 158 \end{pmatrix}$ & $\begin{pmatrix}1080\\ 132\\ 18.8\end{pmatrix}$ & $\begin{pmatrix}644\\ 260\\ 54.3\end{pmatrix}$ & $\begin{pmatrix}287\\ 248\\ 85.8\end{pmatrix}$ & $\begin{pmatrix}23.2\\ 22.9\\ 18.8\end{pmatrix}$\\
\hline 
\hline
Sig. 1 analysis 
0\% (10\%)  &  41.8 (18.3) & 4.34 (1.67) & 276 (256) & -- & --\\
Sig. 2 analysis 
0\% (10\%)  & 130 (72.5) & 15.0 (8.35) & -- & 109 (81.1) & --\\
Sig. 3 analysis  
0\% (10\%)  & 10.5 (10.5) & 1.09 (1.09) & -- & -- & 5.51 (5.51) \\
\hline
\hline
\end{tabular}
\caption{Cut table for $\mathcal{L}=300$ fb$^{-1}$.    In the  fifth row the three entries are for cuts on $H_T$ of 1500, 2250, 3000 GeV, respectively. The final results, shown in the bottom three rows for $\delta=0$ and $\delta = 10\%$, have $H_T>\frac{3}{2}M_B$ and require events to land in the appropriate $(M_{\zp},M_B)$ window, defined in Table \ref{tab:signalwindows}.}
\label{tab:benchmark_results}
\end{table}
\begin{table}
\begin{tabular}[h]{||c|c||c|c|c||}
\hline
\hline
 \multicolumn{2}{||c||}{} & Benchmark Window 1 & Benchmark Window 2 & Benchmark Window 3\\
 \hline
 \hline
 \multicolumn{2}{||c||}{full analysis, with $n_b \ge 3$}& $19.9 \pm 1.1$ & $80.9 \pm 1.7$ & $11.6 \pm 0.3$\\
 \hline
 \multirow{2}{*}{
 $\bm{n_b=0}$ 
 }
 & $n_b \ge 3$ estimate
 & $21.7 \pm 0.4 \pm 0.3$ & $78.5 \pm 0.5 \pm 1.9$ & $12.0 \pm 0.1 \pm 0.7$\\
 & $S/B$ in window &  $ 8.3 \times 10^{-3}$  & $ 3.5 \times 10^{-4}$  & $1.7 \times 10^{-4}$ \\
  \hline
 \multirow{2}{*} {
 $\bm{n_b=1}$ 
 }
 & $n_b \ge 3$ estimate  & $21.9 \pm 0.4 \pm 0.6$ & $79.6 \pm 0.6 \pm 2.4$ & $12.1 \pm 0.1 \pm 0.9$\\
 & $S/B$ in window  &  $0.15$  & $ 6.3 \times 10^{-3}$  & $2.9 \times 10^{-3}$ \\
  \hline
 \multirow{2}{*} {
 $\bm{n_b=2}$ 
 }
 & $n_b \ge 3$ estimate  & $21.2 \pm 0.6 \pm 1.5$ & $79.7 \pm 0.9 \pm 4.0$ & $12.0 \pm 0.2 \pm 1.6$\\
 & $S/B$ in window  &  $1.4$  & $6.4 \times 10^{-2}$  & $2.9 \times 10^{-2}$ \\
 \hline
  \hline
\end{tabular}
\caption{
For $\mathcal{L}=300$ fb$^{-1}$, actual background counts (top row) and the associated estimates using events with zero, one, or two $b$-tagged jets.  The actual  counts come with Monte Carlo uncertainties, and the estimates come with Monte Carlo uncertainties followed by statistical uncertainties associated with the estimation method.  Also shown are signal-to-background ratios for each window and $n_b$ requirement. 
}
\label{tab:background_estimate}
\end{table}

\begin{figure}[t] 
   \centering

   \includegraphics[width=0.32\textwidth]{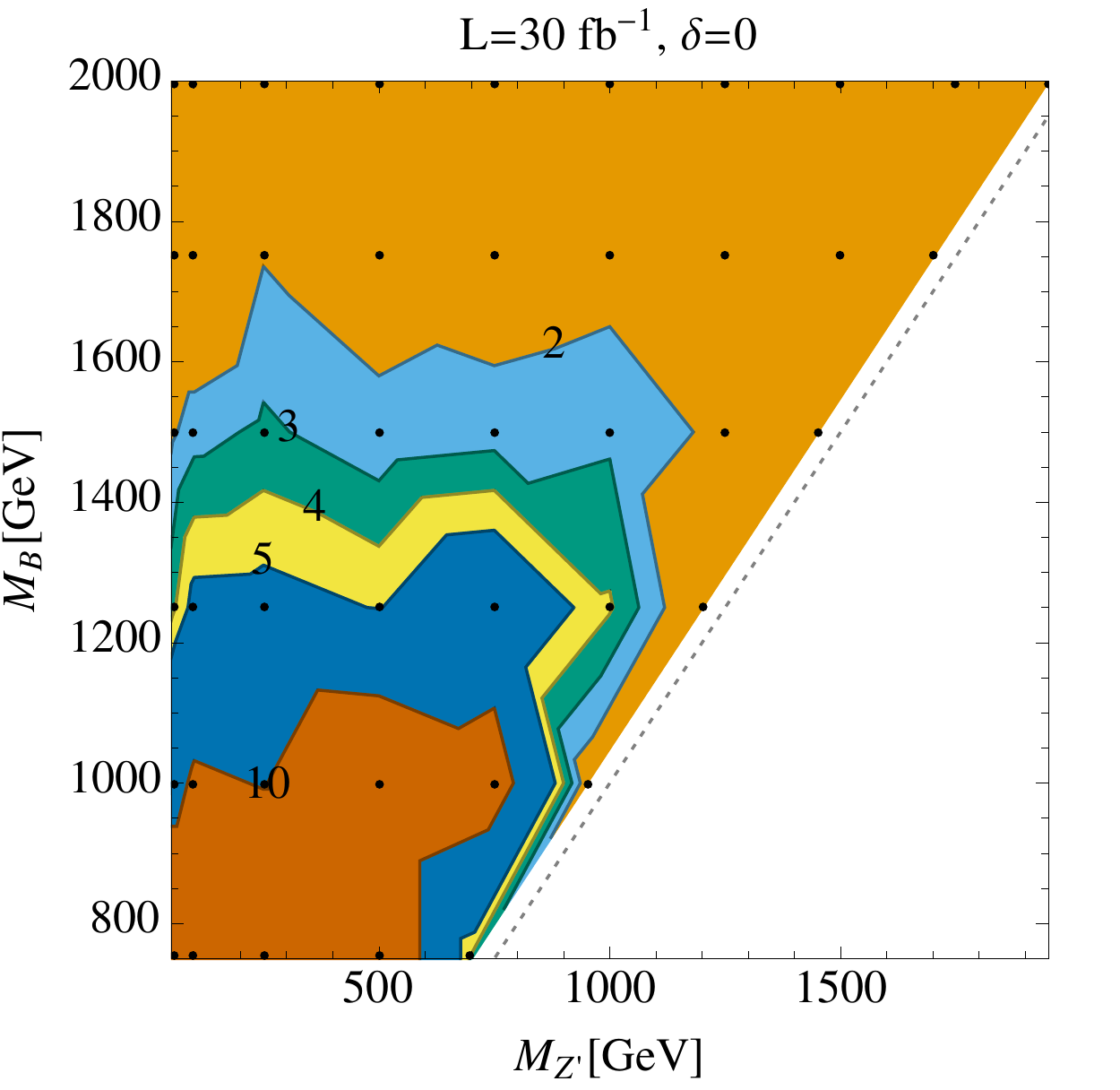} 
   \includegraphics[width=0.32\textwidth]{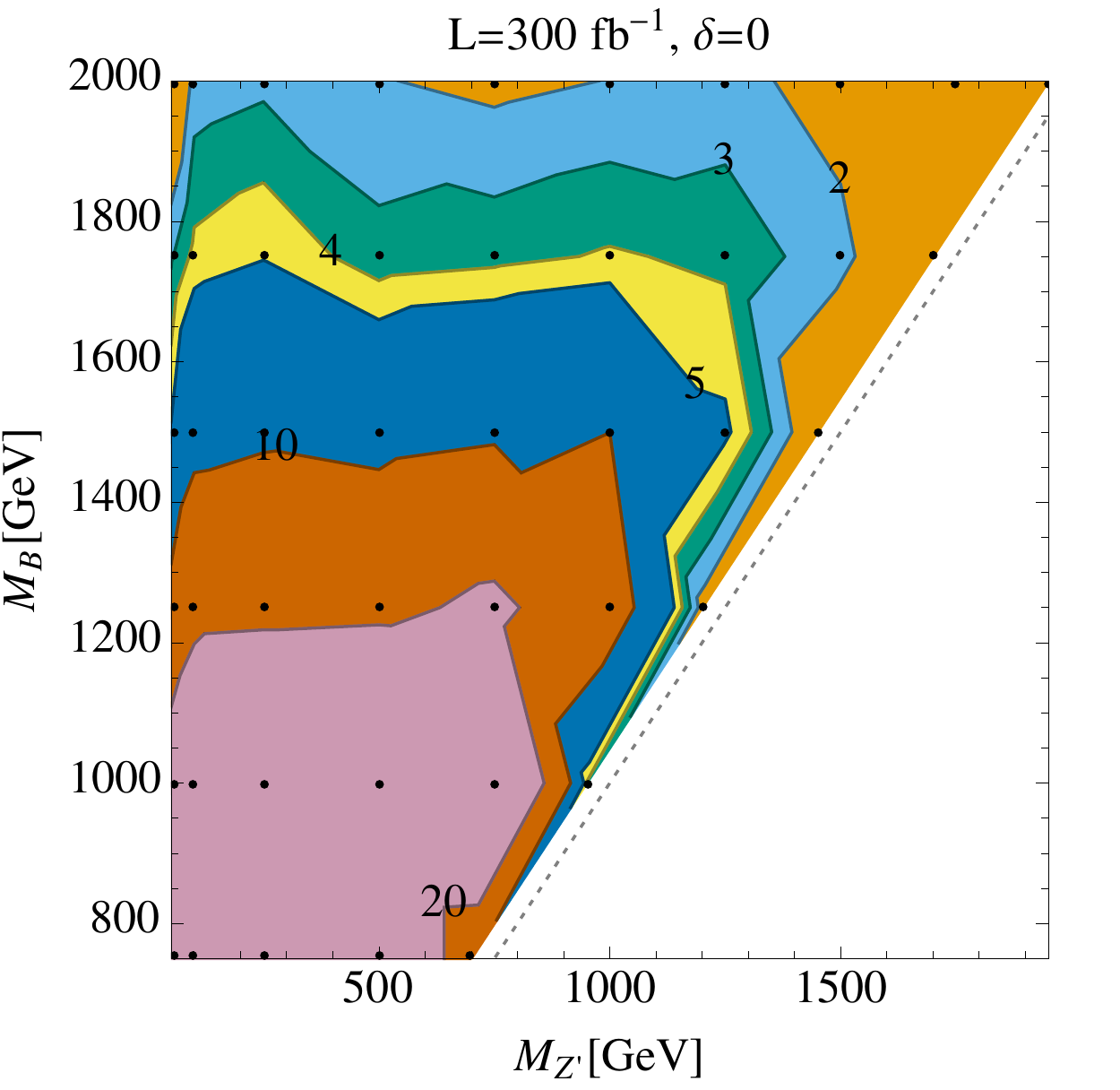}
   \includegraphics[width=0.32\textwidth]{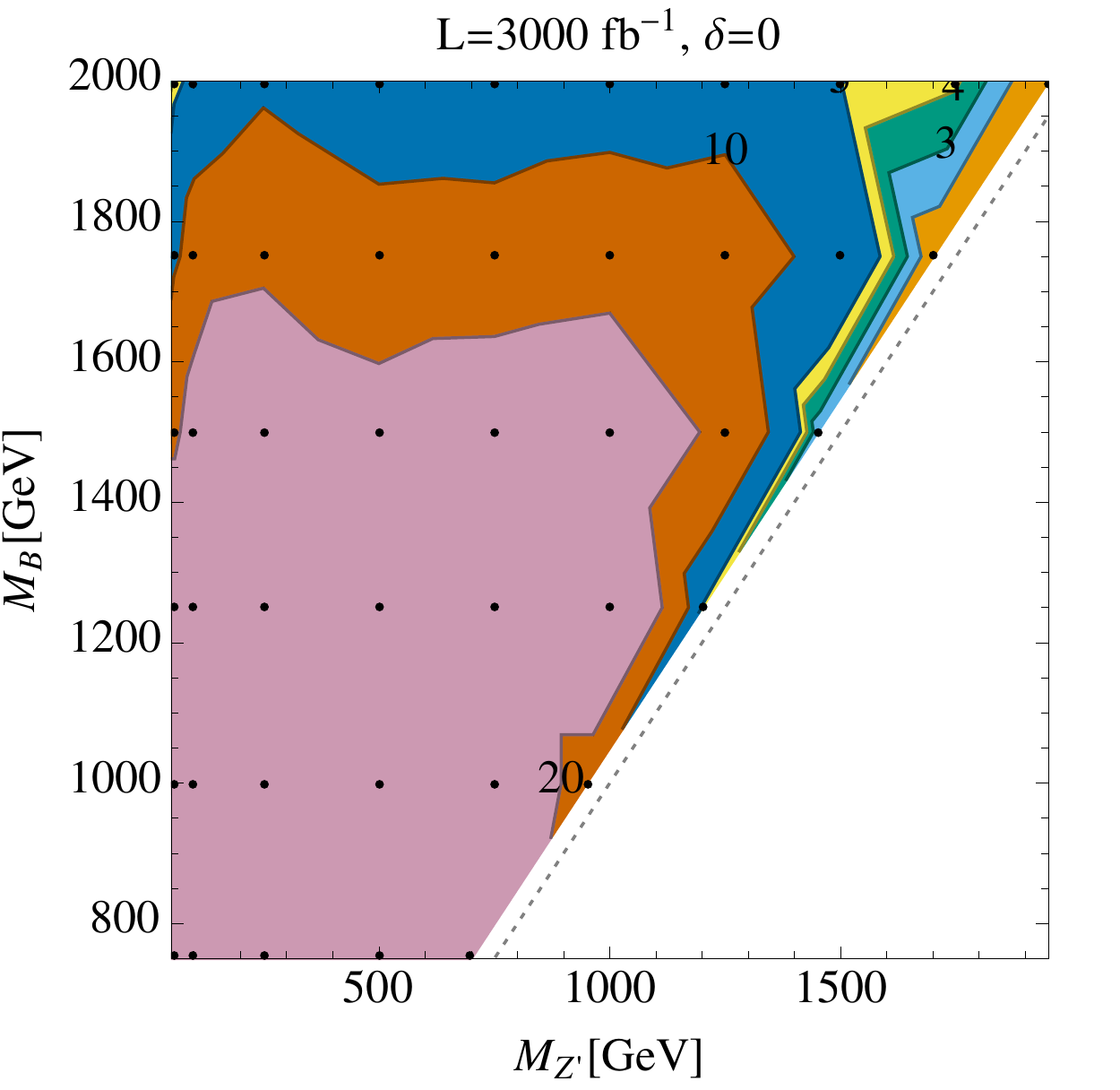}
   \includegraphics[width=0.32\textwidth]{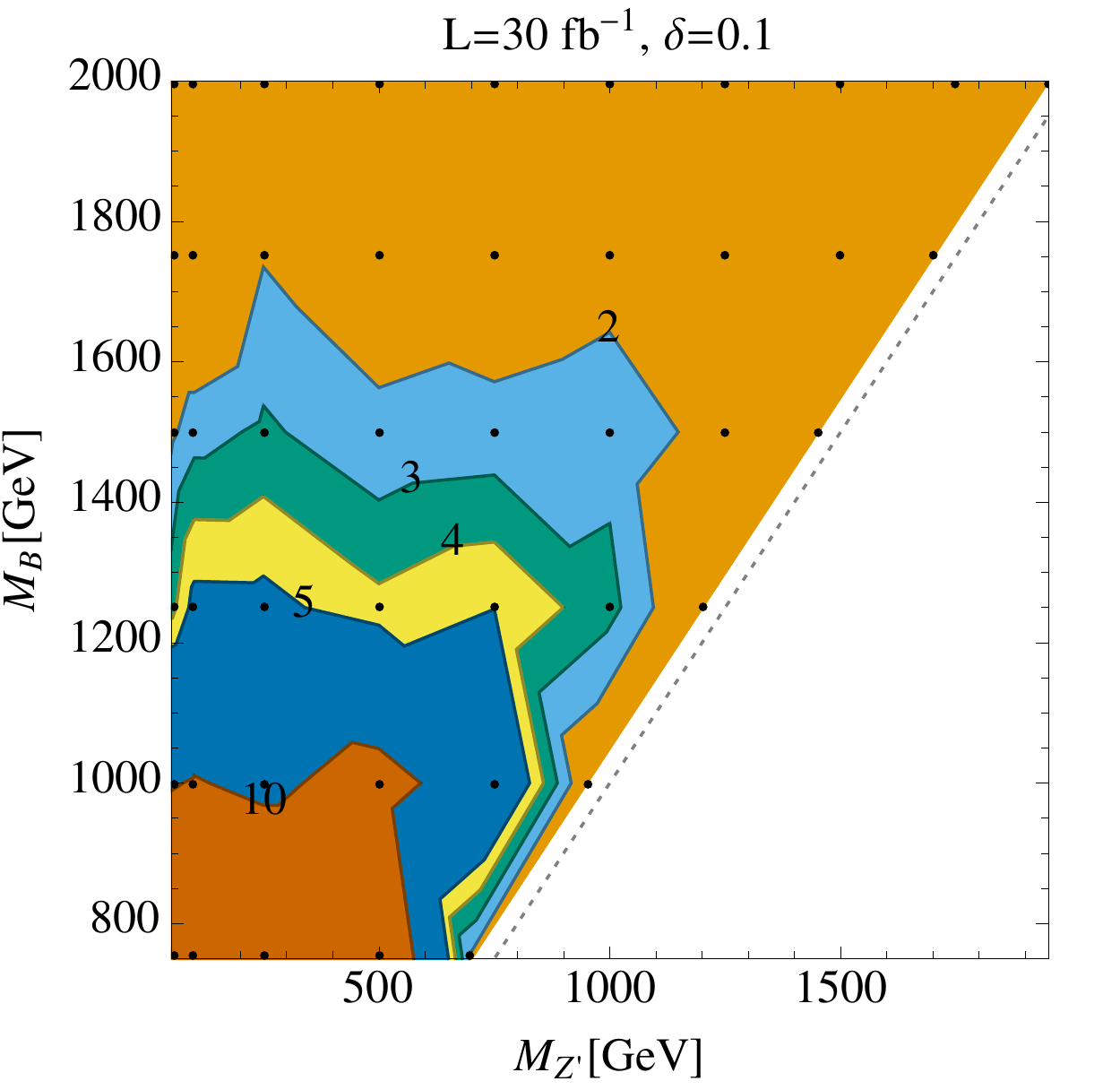} 
   \includegraphics[width=0.32\textwidth]{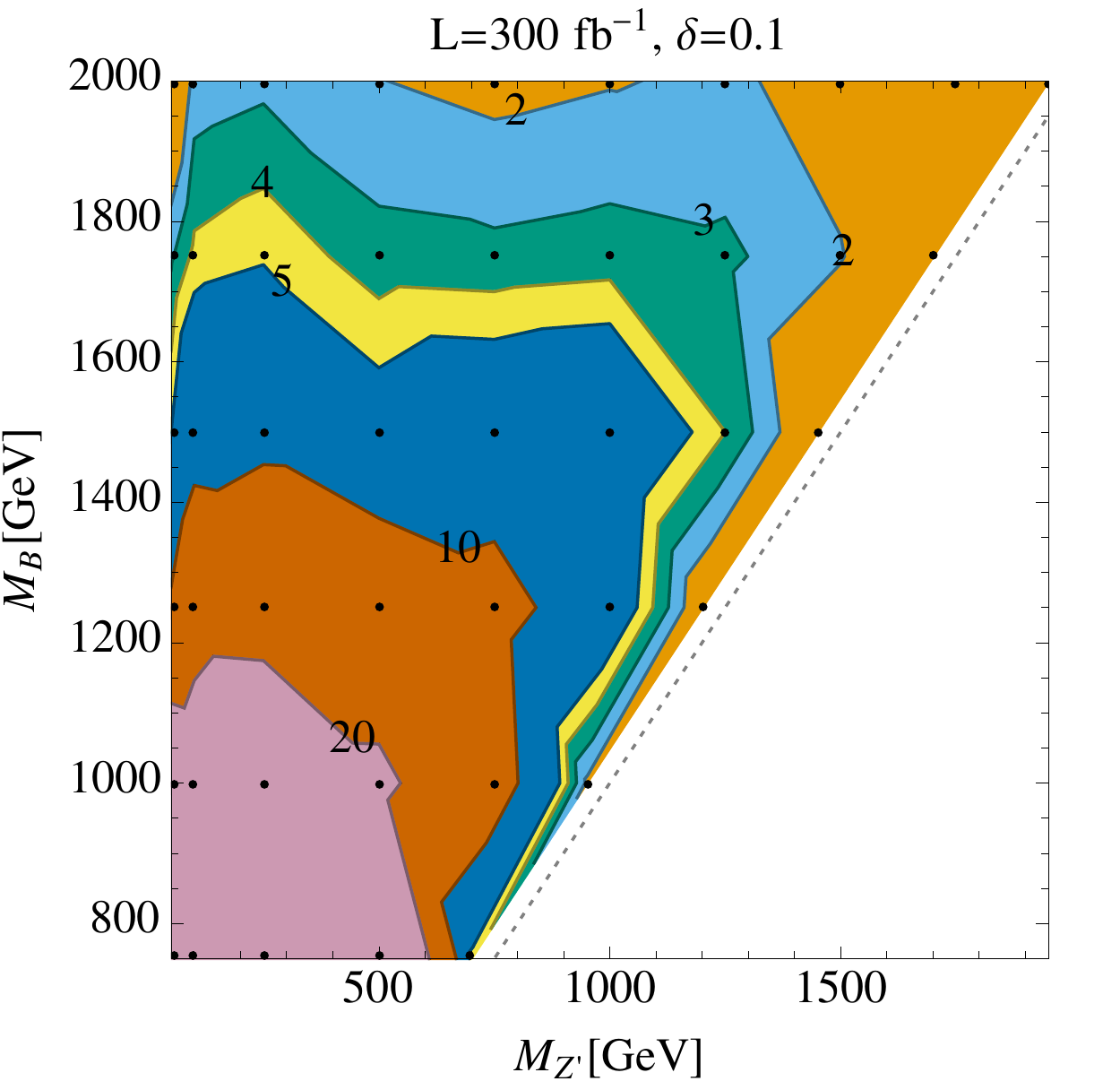}
   \includegraphics[width=0.32\textwidth]{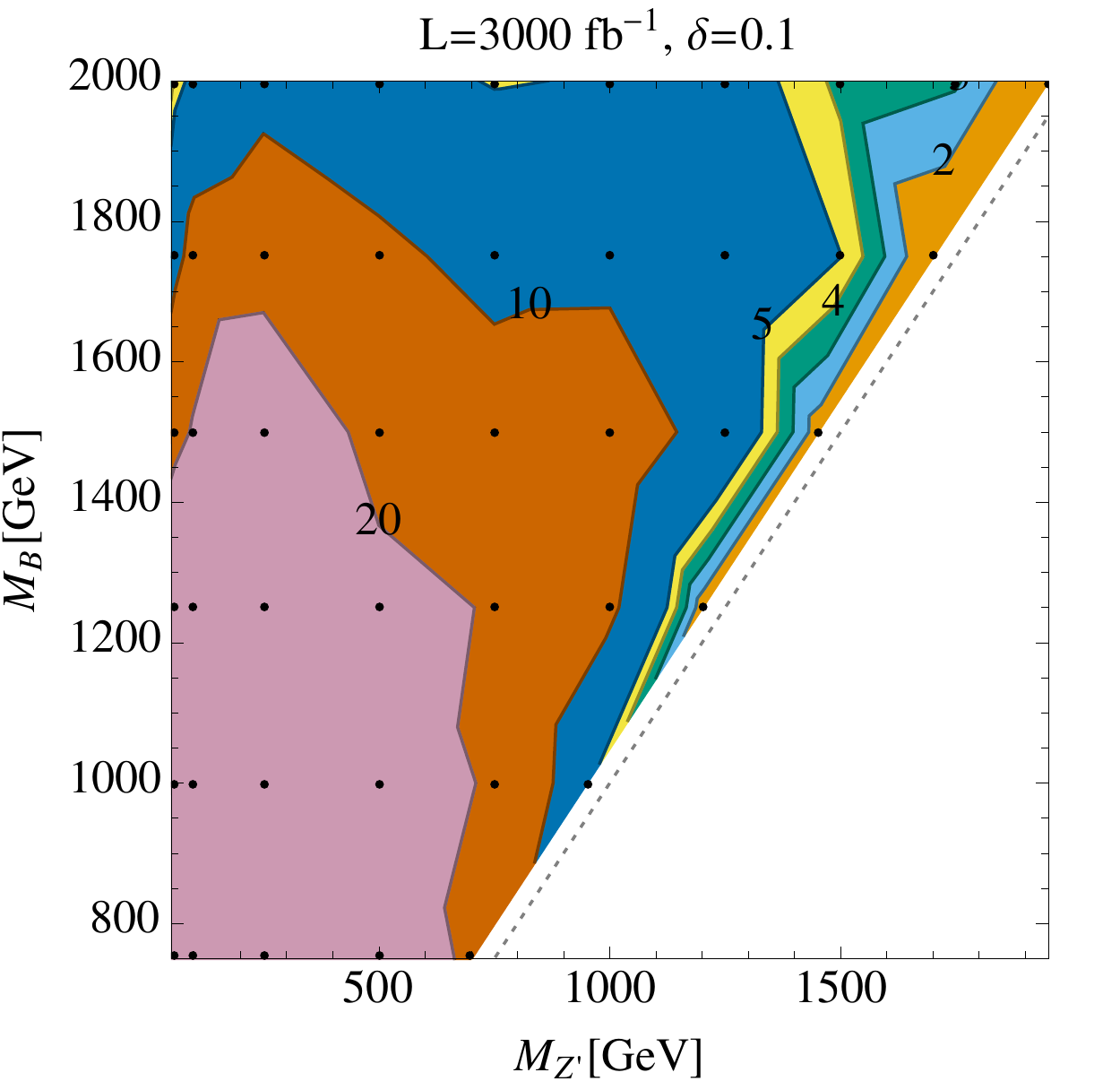}
\caption{
Significance contours for $\mathcal{L}=$ 30, 300, and 3000 fb$^{-1}$, and for $\delta =$ 0 and 10\%.  The black points indicate the signal parameter points  from which the contour plot was 
generated.  
}
   \label{fig:significances}
\end{figure}

Figure~\ref{fig:significances} shows the projected discovery potential in the $M_{Z'} - M_B$ plane for $\mathcal{L}=30$ fb$^{-1}$, 300 fb$^{-1}$, and 3000 fb$^{-1}$.  Discovery at the 5$\sigma$ level is possible for a broad range  of $M_{Z'}$, with $M_B \lsim 1250$ GeV for 30 fb$^{-1}$, with $M_B \lsim 1600$ GeV for 300 fb$^{-1}$, and with $M_B \lsim 2000$ GeV for 3000 fb$^{-1}$.

\section{Conclusions}

The hunt for new colored fermions is an integral part of the broad search strategy employed at the LHC.  To date almost all searches for new vector-like 
partners of the top or bottom quarks have been in final states containing SM bosons ($W,\, Z,\,$ or $h$).  We have pointed out that, by virtue of being vector-like, it is straightforward for the heavy quarks to be charged under additional gauge groups, and that these couplings may dominate their decays.  We have focussed on the simple case of a new $U(1)^\prime$ group which a vector-like $B$ quark is charged under.  We have described a simple realisation of this scenario, based around the concept of the ``Effective $\zp$".  We have outlined the wide range of new phenomena and interesting search channels that exist in this class of simple models, which contain only three new particles.  If the kinetic mixing between $U(1)^\prime$ and hypercharge is small the new channels all involve multiple $b$ quarks.  We demonstrated that there is a broad region of parameter space in these models where the 
new decay $B\rightarrow b \zp/\phi \rightarrow b(b\bar{b})$ dominates.  

We have presented a search method that can simultaneously observe the new quark and the new gauge boson in final states containing up to six $b$ quarks, by carrying out a two-dimensional mass reconstruction of events.  The large QCD and smaller $t\bar{t}$ backgrounds can be effectively reduced by requiring pairs of resonances whose masses are close, which in turn contain sub-resonances whose masses reconstruct to be the same.  Although there are many $b$ quarks in the final state we take a conservative approach and require only three $b$-tags.  A better understanding of $b$-tagging efficiencies may allow this requirement to be strengthened, leading to a further suppression of background.  The kinematics of the process are sensitive to the mass splitting between $B$ and $\zp$ and we account for this be varying our reconstruction technique with the number of final state jets and employing the techniques of $N$-subjettiness to uncover merged jets from the $\zp$ decay.  We find that discovery at the 5$\sigma$ level is possible for a broad range  of $M_{Z'}$, with $M_B \lsim 1250$ GeV for 30 fb$^{-1}$, with $M_B \lsim 1600$ GeV for 300 fb$^{-1}$, and with $M_B \lsim 2000$ GeV for 3000 fb$^{-1}$.  

It is intriguing that the recently observed Galactic center excess can be explained by weak scale DM annihilating into $b$ quarks.  If this takes place through a new mediator one might expect new $b$-quark partners which may themselves decay into the mediator.  We have provided one such example of this and have shown that the LHC has the capability to test this DM scenario over much of its parameter space.

Finally, the technique we describe is not unique to the model we analyse and will be widely applicable to many models where a new particle is pair produced and decays to a lighter new state, finally decaying to SM particles.  For instance, the approach we advocate has an obvious extension to vector-like top quarks, $T\rightarrow t \zp \rightarrow t (b\bar{b})/(t\bar{t})$. 
It would also enhance RPV gluino searches \cite{Chatrchyan:2013gia,Aad:2015lea} in the case where the squarks are lighter than the gluinos.

\subsection*{Note Added} While this work was in the final stages of completion CMS released details of a search for $T$ in the exotic mode $T\rightarrow b W^\prime$ with $W'$ decaying leptonically \cite{CMS-PAS-B2G-12-025}.  The CMS analysis also searches simultaneously for two new particles and carries out a two-dimensional mass reconstruction of events, but the final state and particle content are different from what we consider.

\section*{Acknowledgements}
We would like to thank John Campbell for helpful conversations.  We would like to thank the Aspen Center for Physics, where this work was initiated, for their hospitality.  Aspen Center for Physics is supported by the National Science Foundation Grant No. PHY-1066293.  Fermilab is operated by Fermi Research Alliance, LLC under Contract No. DE-AC02-07CH11359 with the United States Department of Energy.  The work of DTS was supported by NSF Grant \#1216168.  

\bibliographystyle{JHEP}
\bibliography{Bfriendly}

\providecommand{\href}[2]{#2}\begingroup\raggedright\begin{thebibliography}{10}

\bibitem{Khachatryan:2015gza}
{\bf CMS} Collaboration, V.~Khachatryan {\em et.~al.}, {\it {Search for
  pair-produced vector-like B quarks in proton-proton collisions at $\sqrt{s}$
  = 8 TeV}},  \href{http://xxx.lanl.gov/abs/1507.0712}{{\tt arXiv:1507.0712}}.

\bibitem{Aad:2015mba}
{\bf ATLAS} Collaboration, G.~Aad {\em et.~al.}, {\it {Search for vector-like
  $B$ quarks in events with one isolated lepton, missing transverse momentum
  and jets at $\sqrt{s}=$ 8 TeV with the ATLAS detector}},
  \href{http://xxx.lanl.gov/abs/1503.0542}{{\tt arXiv:1503.0542}}.

\bibitem{Aad:2014efa}
{\bf ATLAS} Collaboration, G.~Aad {\em et.~al.}, {\it {Search for pair and
  single production of new heavy quarks that decay to a $Z$ boson and a
  third-generation quark in $pp$ collisions at $\sqrt{s}=8$ TeV with the ATLAS
  detector}},  {\em JHEP} {\bf 1411} (2014) 104,
  [\href{http://xxx.lanl.gov/abs/1409.5500}{{\tt arXiv:1409.5500}}].

\bibitem{Chatrchyan:2013uxa}
{\bf CMS} Collaboration, S.~Chatrchyan {\em et.~al.}, {\it {Inclusive search
  for a vector-like T quark with charge $\frac{2}{3}$ in pp collisions at
  $\sqrt{s}$ = 8 TeV}},  {\em Phys.Lett.} {\bf B729} (2014) 149--171,
  [\href{http://xxx.lanl.gov/abs/1311.7667}{{\tt arXiv:1311.7667}}].

\bibitem{CMS-PAS-B2G-12-017}
{\bf CMS Collaboration} Collaboration, {\it {Search for vector-like quarks in
  final states with a single lepton and jets in pp collisions at sqrt s = 8
  TeV}},  Tech. Rep. CMS-PAS-B2G-12-017, CERN, Geneva, 2014.

\bibitem{TheATLAScollaboration:2013sha}
{\bf ATLAS} Collaboration, T.~A. collaboration, {\it {Search for pair
  production of heavy top-like quarks decaying to a high-$p_{\rm T}$ $W$ boson
  and a $b$ quark in the lepton plus jets final state in $pp$ collisions at
  $\sqrt{s}=8$ TeV with the ATLAS detector}}, .

\bibitem{ATLAS:2013ima}
{\bf ATLAS} Collaboration, {\it {Search for heavy top-like quarks decaying to a
  Higgs boson and a top quark in the lepton plus jets final state in $pp$
  collisions at $\sqrt{s}=8$ TeV with the ATLAS detector}}, .

\bibitem{Fox:2011qd}
P.~J. Fox, J.~Liu, D.~Tucker-Smith, and N.~Weiner, {\it {An Effective Z'}},
  {\em Phys.Rev.} {\bf D84} (2011) 115006,
  [\href{http://xxx.lanl.gov/abs/1104.4127}{{\tt arXiv:1104.4127}}].

\bibitem{Goodenough:2009gk}
L.~Goodenough and D.~Hooper, {\it {Possible Evidence For Dark Matter
  Annihilation In The Inner Milky Way From The Fermi Gamma Ray Space
  Telescope}},  \href{http://xxx.lanl.gov/abs/0910.2998}{{\tt
  arXiv:0910.2998}}.

\bibitem{Hooper:2010mq}
D.~Hooper and L.~Goodenough, {\it {Dark Matter Annihilation in the Galactic
  Center as Seen by the Fermi Gamma Ray Space Telescope}},
  \href{http://xxx.lanl.gov/abs/1010.2752}{{\tt arXiv:1010.2752}}.

\bibitem{Thaler:2010tr}
J.~Thaler and K.~Van~Tilburg, {\it {Identifying Boosted Objects with
  N-subjettiness}},  {\em JHEP} {\bf 1103} (2011) 015,
  [\href{http://xxx.lanl.gov/abs/1011.2268}{{\tt arXiv:1011.2268}}].

\bibitem{Jackson:2013rqp}
C.~B. Jackson, G.~Servant, G.~Shaughnessy, T.~M.~P. Tait, and M.~Taoso, {\it
  {Gamma Rays from Top-Mediated Dark Matter Annihilations}},  {\em JCAP} {\bf
  1307} (2013) 006, [\href{http://xxx.lanl.gov/abs/1303.4717}{{\tt
  arXiv:1303.4717}}].

\bibitem{Curtin:2013fra}
D.~Curtin {\em et.~al.}, {\it {Exotic decays of the 125 GeV Higgs boson}},
  {\em Phys. Rev.} {\bf D90} (2014), no.~7 075004,
  [\href{http://xxx.lanl.gov/abs/1312.4992}{{\tt arXiv:1312.4992}}].

\bibitem{Khachatryan:2014fba}
{\bf CMS} Collaboration, V.~Khachatryan {\em et.~al.}, {\it {Search for physics
  beyond the standard model in dilepton mass spectra in proton-proton
  collisions at $ \sqrt{s}=8 $ TeV}},  {\em JHEP} {\bf 1504} (2015) 025,
  [\href{http://xxx.lanl.gov/abs/1412.6302}{{\tt arXiv:1412.6302}}].

\bibitem{Boyarsky:2010dr}
A.~Boyarsky, D.~Malyshev, and O.~Ruchayskiy, {\it {A Comment on the Emission
  from the Galactic Center as Seen by the Fermi Telescope}},  {\em Phys.Lett.}
  {\bf B705} (2011) 165--169, [\href{http://xxx.lanl.gov/abs/1012.5839}{{\tt
  arXiv:1012.5839}}].

\bibitem{Hooper:2011ti}
D.~Hooper and T.~Linden, {\it {On the Origin of the Gamma Rays from the
  Galactic Center}},  {\em Phys.Rev.} {\bf D84} (2011) 123005,
  [\href{http://xxx.lanl.gov/abs/1110.0006}{{\tt arXiv:1110.0006}}].

\bibitem{Linden:2012iv}
T.~Linden, E.~Lovegrove, and S.~Profumo, {\it {The Morphology of Hadronic
  Emission Models for the Gamma-Ray Source at the Galactic Center}},  {\em
  Astrophys.J.} {\bf 753} (2012) 41,
  [\href{http://xxx.lanl.gov/abs/1203.3539}{{\tt arXiv:1203.3539}}].

\bibitem{Abazajian:2012pn}
K.~N. Abazajian and M.~Kaplinghat, {\it {Detection of a Gamma-Ray Source in the
  Galactic Center Consistent with Extended Emission from Dark Matter
  Annihilation and Concentrated Astrophysical Emission}},  {\em Phys.Rev.} {\bf
  D86} (2012) 083511, [\href{http://xxx.lanl.gov/abs/1207.6047}{{\tt
  arXiv:1207.6047}}].

\bibitem{Hooper:2013rwa}
D.~Hooper and T.~R. Slatyer, {\it {Two Emission Mechanisms in the Fermi
  Bubbles: a Possible Signal of Annihilating Dark Matter}},  {\em Phys.Dark
  Univ.} {\bf 2} (2013) 118--138,
  [\href{http://xxx.lanl.gov/abs/1302.6589}{{\tt arXiv:1302.6589}}].

\bibitem{Gordon:2013vta}
C.~Gordon and O.~Macias, {\it {Dark Matter and Pulsar Model Constraints from
  Galactic Center Fermi-Lat Gamma Ray Observations}},  {\em Phys.Rev.} {\bf
  D88} (2013), no.~8 083521, [\href{http://xxx.lanl.gov/abs/1306.5725}{{\tt
  arXiv:1306.5725}}].

\bibitem{Abazajian:2014fta}
K.~N. Abazajian, N.~Canac, S.~Horiuchi, and M.~Kaplinghat, {\it {Astrophysical
  and Dark Matter Interpretations of Extended Gamma-Ray Emission from the
  Galactic Center}},  {\em Phys.Rev.} {\bf D90} (2014), no.~2 023526,
  [\href{http://xxx.lanl.gov/abs/1402.4090}{{\tt arXiv:1402.4090}}].

\bibitem{Daylan:2014rsa}
T.~Daylan, D.~P. Finkbeiner, D.~Hooper, T.~Linden, S.~K.~N. Portillo, {\em
  et.~al.}, {\it {The Characterization of the Gamma-Ray Signal from the Central
  Milky Way: a Compelling Case for Annihilating Dark Matter}},
  \href{http://xxx.lanl.gov/abs/1402.6703}{{\tt arXiv:1402.6703}}.

\bibitem{Zhou:2014lva}
B.~Zhou, Y.-F. Liang, X.~Huang, X.~Li, Y.-Z. Fan, {\em et.~al.}, {\it {GeV
  Excess in the Milky Way: the Role of Diffuse Galactic Gamma Ray Emission
  Template}},  \href{http://xxx.lanl.gov/abs/1406.6948}{{\tt arXiv:1406.6948}}.

\bibitem{Calore:2014xka}
F.~Calore, I.~Cholis, and C.~Weniger, {\it {Background Model Systematics for
  the Fermi GeV Excess}},  {\em JCAP} {\bf 1503} (2015) 038,
  [\href{http://xxx.lanl.gov/abs/1409.0042}{{\tt arXiv:1409.0042}}].

\bibitem{Agrawal:2014oha}
P.~Agrawal, B.~Batell, P.~J. Fox, and R.~Harnik, {\it {Wimps at the Galactic
  Center}},  {\em JCAP} {\bf 1505} (2015), no.~05 011,
  [\href{http://xxx.lanl.gov/abs/1411.2592}{{\tt arXiv:1411.2592}}].

\bibitem{Boehm:2014bia}
C.~Boehm, M.~J. Dolan, and C.~McCabe, {\it {A Weighty Interpretation of the
  Galactic Centre Excess}},  {\em Phys.Rev.} {\bf D90} (2014), no.~2 023531,
  [\href{http://xxx.lanl.gov/abs/1404.4977}{{\tt arXiv:1404.4977}}].

\bibitem{Ko:2014gha}
P.~Ko, W.-I. Park, and Y.~Tang, {\it {Higgs Portal Vector Dark Matter for
  $\mathinner{\mathrm{GeV}}$ Scale $\gamma$-ray Excess from Galactic Center}},
  {\em JCAP} {\bf 1409} (2014) 013,
  [\href{http://xxx.lanl.gov/abs/1404.5257}{{\tt arXiv:1404.5257}}].

\bibitem{Abdullah:2014lla}
M.~Abdullah, A.~DiFranzo, A.~Rajaraman, T.~M. Tait, P.~Tanedo, {\em et.~al.},
  {\it {Hidden On-Shell Mediators for the Galactic Center $\gamma$-ray
  Excess}},  {\em Phys.Rev.} {\bf D90} (2014), no.~3 035004,
  [\href{http://xxx.lanl.gov/abs/1404.6528}{{\tt arXiv:1404.6528}}].

\bibitem{Martin:2014sxa}
A.~Martin, J.~Shelton, and J.~Unwin, {\it {Fitting the Galactic Center
  Gamma-Ray Excess with Cascade Annihilations}},  {\em Phys.Rev.} {\bf D90}
  (2014), no.~10 103513, [\href{http://xxx.lanl.gov/abs/1405.0272}{{\tt
  arXiv:1405.0272}}].

\bibitem{Pospelov:2007mp}
M.~Pospelov, A.~Ritz, and M.~B. Voloshin, {\it {Secluded WIMP Dark Matter}},
  {\em Phys.Lett.} {\bf B662} (2008) 53--61,
  [\href{http://xxx.lanl.gov/abs/0711.4866}{{\tt arXiv:0711.4866}}].

\bibitem{Belanger:2013oya}
G.~Belanger, F.~Boudjema, A.~Pukhov, and A.~Semenov, {\it {micrOMEGAs3: A
  program for calculating dark matter observables}},  {\em Comput. Phys.
  Commun.} {\bf 185} (2014) 960--985,
  [\href{http://xxx.lanl.gov/abs/1305.0237}{{\tt arXiv:1305.0237}}].

\bibitem{Aad:2013ija}
{\bf ATLAS} Collaboration, G.~Aad {\em et.~al.}, {\it {Search for direct
  third-generation squark pair production in final states with missing
  transverse momentum and two $b$-jets in $\sqrt{s} =$ 8 TeV $pp$ collisions
  with the ATLAS detector}},  {\em JHEP} {\bf 10} (2013) 189,
  [\href{http://xxx.lanl.gov/abs/1308.2631}{{\tt arXiv:1308.2631}}].

\bibitem{Khachatryan:2015wza}
{\bf CMS} Collaboration, V.~Khachatryan {\em et.~al.}, {\it {Searches for
  third-generation squark production in fully hadronic final states in
  proton-proton collisions at $ \sqrt{s} = 8$ TeV}},  {\em JHEP} {\bf 06}
  (2015) 116, [\href{http://xxx.lanl.gov/abs/1503.0803}{{\tt
  arXiv:1503.0803}}].

\bibitem{Akerib:2013tjd}
{\bf LUX} Collaboration, D.~S. Akerib {\em et.~al.}, {\it {First results from
  the LUX dark matter experiment at the Sanford Underground Research
  Facility}},  {\em Phys. Rev. Lett.} {\bf 112} (2014) 091303,
  [\href{http://xxx.lanl.gov/abs/1310.8214}{{\tt arXiv:1310.8214}}].

\bibitem{CMS:2013una}
{\bf CMS} Collaboration, C.~Collaboration, {\it {Search for Vector-Like b' Pair
  Production with Multilepton Final States in pp collisions at sqrt(s) = 8
  TeV}}, .

\bibitem{atlasVLQ}
{\bf ATLAS} Collaboration, {The ATLAS collaboration}, {\it {Search for
  production of vector-like quark pairs and of four top quarks in the lepton
  plus jets final state in $pp$ collisions at $\sqrt{s}=8$ TeV with the ATLAS
  detector}}, .

\bibitem{Khachatryan:2015tra}
{\bf CMS} Collaboration, V.~Khachatryan {\em et.~al.}, {\it {Search for Neutral
  MSSM Higgs Bosons Decaying into A Pair of Bottom Quarks}},
  \href{http://xxx.lanl.gov/abs/1506.0832}{{\tt arXiv:1506.0832}}.

\bibitem{Alloul:2013bka}
A.~Alloul, N.~D. Christensen, C.~Degrande, C.~Duhr, and B.~Fuks, {\it
  {FeynRules 2.0 - A complete toolbox for tree-level phenomenology}},  {\em
  Comput. Phys. Commun.} {\bf 185} (2014) 2250--2300,
  [\href{http://xxx.lanl.gov/abs/1310.1921}{{\tt arXiv:1310.1921}}].

\bibitem{Alwall:2014hca}
J.~Alwall, R.~Frederix, S.~Frixione, V.~Hirschi, F.~Maltoni, {\em et.~al.},
  {\it {The automated computation of tree-level and next-to-leading order
  differential cross sections, and their matching to parton shower
  simulations}},  {\em JHEP} {\bf 1407} (2014) 079,
  [\href{http://xxx.lanl.gov/abs/1405.0301}{{\tt arXiv:1405.0301}}].

\bibitem{Sjostrand:2007gs}
T.~Sjostrand, S.~Mrenna, and P.~Z. Skands, {\it {A Brief Introduction to PYTHIA
  8.1}},  {\em Comput.Phys.Commun.} {\bf 178} (2008) 852--867,
  [\href{http://xxx.lanl.gov/abs/0710.3820}{{\tt arXiv:0710.3820}}].

\bibitem{deFavereau:2013fsa}
{\bf DELPHES 3} Collaboration, J.~de~Favereau {\em et.~al.}, {\it {DELPHES 3, A
  modular framework for fast simulation of a generic collider experiment}},
  {\em JHEP} {\bf 1402} (2014) 057,
  [\href{http://xxx.lanl.gov/abs/1307.6346}{{\tt arXiv:1307.6346}}].

\bibitem{Cacciari:2011ma}
M.~Cacciari, G.~P. Salam, and G.~Soyez, {\it {FastJet User Manual}},  {\em
  Eur.Phys.J.} {\bf C72} (2012) 1896,
  [\href{http://xxx.lanl.gov/abs/1111.6097}{{\tt arXiv:1111.6097}}].

\bibitem{Cacciari:2008gp}
M.~Cacciari, G.~P. Salam, and G.~Soyez, {\it {The Anti-k(t) jet clustering
  algorithm}},  {\em JHEP} {\bf 0804} (2008) 063,
  [\href{http://xxx.lanl.gov/abs/0802.1189}{{\tt arXiv:0802.1189}}].

\bibitem{Aliev:2010zk}
M.~Aliev, H.~Lacker, U.~Langenfeld, S.~Moch, P.~Uwer, {\em et.~al.}, {\it
  {HATHOR: HAdronic Top and Heavy quarks crOss section calculatoR}},  {\em
  Comput.Phys.Commun.} {\bf 182} (2011) 1034--1046,
  [\href{http://xxx.lanl.gov/abs/1007.1327}{{\tt arXiv:1007.1327}}].

\bibitem{Czakon:2013goa}
M.~Czakon, P.~Fiedler, and A.~Mitov, {\it {Total Top-Quark Pair-Production
  Cross Section at Hadron Colliders Through $O(\alpha_S^4)$}},  {\em
  Phys.Rev.Lett.} {\bf 110} (2013) 252004,
  [\href{http://xxx.lanl.gov/abs/1303.6254}{{\tt arXiv:1303.6254}}].

\bibitem{Martin:2009iq}
A.~Martin, W.~Stirling, R.~Thorne, and G.~Watt, {\it {Parton distributions for
  the LHC}},  {\em Eur.Phys.J.} {\bf C63} (2009) 189--285,
  [\href{http://xxx.lanl.gov/abs/0901.0002}{{\tt arXiv:0901.0002}}].

\bibitem{Aad:2011tqa}
{\bf ATLAS} Collaboration, G.~Aad {\em et.~al.}, {\it {Measurement of multi-jet
  cross sections in proton-proton collisions at a 7 TeV center-of-mass
  energy}},  {\em Eur.Phys.J.} {\bf C71} (2011) 1763,
  [\href{http://xxx.lanl.gov/abs/1107.2092}{{\tt arXiv:1107.2092}}].

\bibitem{Karneyeu:2013aha}
A.~Karneyeu, L.~Mijovic, S.~Prestel, and P.~Skands, {\it {MCPLOTS: a particle
  physics resource based on volunteer computing}},  {\em Eur.Phys.J.} {\bf C74}
  (2014) 2714, [\href{http://xxx.lanl.gov/abs/1306.3436}{{\tt
  arXiv:1306.3436}}].

\bibitem{Evans:2014gfa}
J.~A. Evans, {\it {A Swarm of Bs}},  {\em JHEP} {\bf 08} (2014) 073,
  [\href{http://xxx.lanl.gov/abs/1402.4481}{{\tt arXiv:1402.4481}}].

\bibitem{Chatrchyan:2013gia}
{\bf CMS} Collaboration, S.~Chatrchyan {\em et.~al.}, {\it {Searches for light-
  and heavy-flavour three-jet resonances in pp collisions at $\sqrt{s} = 8$
  TeV}},  {\em Phys. Lett.} {\bf B730} (2014) 193--214,
  [\href{http://xxx.lanl.gov/abs/1311.1799}{{\tt arXiv:1311.1799}}].

\bibitem{Aad:2015lea}
{\bf ATLAS} Collaboration, G.~Aad {\em et.~al.}, {\it {Search for massive
  supersymmetric particles decaying to many jets using the ATLAS detector in
  $pp$ collisions at $\sqrt{s} = 8$ TeV}},  {\em Phys. Rev.} {\bf D91} (2015),
  no.~11 112016, [\href{http://xxx.lanl.gov/abs/1502.0568}{{\tt
  arXiv:1502.0568}}].

\bibitem{CMS-PAS-B2G-12-025}
{\bf CMS Collaboration} Collaboration, {\it { Search in two-dimensional mass
  space for $T^\prime T^\prime$ to $W^\prime b W^\prime b$ in the dilepton
  final state in proton-proton collisions at 8 TeV}},  Tech. Rep.
  CMS-PAS-B2G-12-025, CERN, Geneva, 2015.

\end{thebibliography}\endgroup

\end{document}